\expandafter\edef\csname hypers\string @fe\endcsname{\catcode
                                             `\noexpand @=\the\catcode`\@}%
\catcode`\@=11
%
%
\ifx\hyper@utoprocess\hyper@ndefined
\else
 \expandafter\hyper@utoprocess\fi
\ifx\hyperd@ne\hyper@ndefined
 \global\let\hyperd@ne=\relax
\else
 \hypers@fe
 \errhelp{hyperbasics.tex needs to be included only once outside
          of any {...} or \begingroup...\endgroup. You have tried to
          include it more than once. If the previous include was indeed
          outside any groupings, continue and all will be well.}%
 \errmessage{Input this file only once!}%
 \expandafter \fi
%
%
\def\hyperv@rsion{12}%
%
%
\newread\hyperf@le
\def\hyperf@lename{\jobname.hrf}%
\immediate\openin\hyperf@le\hyperf@lename\relax
\ifeof\hyperf@le\relax
 \immediate\closein\hyperf@le\relax
\else
 \immediate\closein\hyperf@le\relax
 \input \hyperf@lename
\fi
%
%
\newwrite\hyperf@le
\immediate\openout\hyperf@le\hyperf@lename
%
%
\newtoks\hypert@ks
%
%
\edef\hypert@mp{\catcode`\noexpand\%=\the\catcode`\%}
\catcode`\%=12
\def\hyperp@rcent{
\hypert@mp
\edef\hypert@mp{\catcode`\noexpand\#=\the\catcode`\#}%
\catcode`\#=12
\def\hyperh@sh{#}%
\hypert@mp
\let\hypert@mp=\relax
\let\hyper@nd=\relax
\def\hyperbl@nk{ }
\def\hyperstr@pquote#1"#2\hyper@nd{
   #1
   \ifx\hyper@nd#2\hyper@nd
   \else\hyperp@rcent22\hyperstr@pquote#2\hyper@nd\fi}%
\def\hyperstr@pblank#1 #2\hyper@nd{
   #1
   \ifx\hyper@nd#2\hyper@nd
   \else\hyperp@rcent20\hyperstr@pblank#2\hyper@nd\fi}
\long\def\hyper@nchor#1#2{\edef\hyperm@cro{html:<A #1>}%
                          \special\expandafter{\hyperm@cro}%
                          {#2}}%
\def\hyper@atm@ning#1->#2\hyper@nd{#2}
\def\hyperlink{\protect\hyperlink@}
\def\hyperlink@{{\catcode\expandafter`\noexpand\#=12 
                 \catcode\expandafter`\noexpand\%=12 
                 \catcode\expandafter`\noexpand\~=12
                \expandafter}\hyperlink@@}
\def\hyperlink@@#1{\protect\hyperlink@@@{#1}}
\def\hyperlink@@@#1{\edef\hypert@mp{#1}%
               \edef\hypert@mp{\expandafter\hyper@atm@ning\meaning\hypert@mp
                               \hyper@nd}%
               \edef\hypert@mp{\expandafter\hyperstr@pquote%
                               \hypert@mp"\hyper@nd}%
               \edef\hypert@mp{\expandafter\expandafter\expandafter
                               \hyperstr@pblank\expandafter%
                               \hypert@mp\hyperbl@nk\hyper@nd}%
               \hyper@nchor{href=\expandafter"\hypert@mp"}}%
\def\hypertarget#1{\edef\hypert@mp{#1}%
               \edef\hypert@mp{\expandafter\hyper@atm@ning\meaning\hypert@mp
                               \hyper@nd}%
               \edef\hypert@mp{\expandafter\hyperstr@pquote%
                               \hypert@mp"\hyper@nd}%
               \edef\hypert@mp{\expandafter\expandafter\expandafter
                               \hyperstr@pblank\expandafter%
                               \hypert@mp\hyperbl@nk\hyper@nd}%
               \hyper@nchor{name=\expandafter"\hypert@mp"}}%
\def\hyperref{\afterassignment\hyperr@f\let\hyperp@ram}
\def\hyperr@f{\ifx\hyperp@ram{\iffalse}\fi
               \expandafter\expandafter\expandafter\hyperr@@
               \expandafter{%
              \else
               \iffalse}\fi
               \ifx\hyperp@ram\hyper@ndefined
                 \message{Undefined reference}%
                 \def\hyperp@r@m{{}{undefined}{}}%
               \else
                 \edef\hyperp@r@m{\hyperp@ram}%
               \fi
               \expandafter\expandafter\expandafter\hyperr@@
               \expandafter\hyperp@r@m
              \fi}%
\def\hyperr@@#1#2#3{\ifx\hyper@ndefined#1\hyper@ndefined
                    \hypert@ks\expandafter{\hyperh@sh#2.#3}%
                    \else
                     \ifx\hyper@ndefined#2#3\hyper@ndefined
                      \hypert@ks{#1}%
                     \else
                      \def\hypert@mp{#1}%
                      \hypert@ks\expandafter\expandafter\expandafter
                      {\expandafter\hypert@mp\hyperh@sh#2.#3}%
                     \fi
                    \fi
                    \expandafter\hyperlink\expandafter{\the\hypert@ks}}%
\def\hyperdef#1#2#3{{\escapechar=`\\\relax
                     \edef\hyper@t@mp@{\hyperstr@pquote#2.#3"\hyper@nd}%
                     \expandafter\ifx\csname hyperd@\meaning\hyper@t@mp@
                     \endcsname
                     \relax
                     \expandafter\gdef\csname hyperd@\meaning\hyper@t@mp@
                     \endcsname{}%
                     \gdef#1{{}{\hyperstr@pquote#2"\hyper@nd}%
                               {\hyperstr@pquote#3"\hyper@nd}}%
                     \immediate\write\hyperf@le{\def\noexpand#1{{}%
                        {\noexpand\hyperstr@pquote#2"\noexpand\hyper@nd}%
                        {\noexpand\hyperstr@pquote#3"\noexpand\hyper@nd}}}%
                     \xdef\hyper@t@mp@{\global\let\noexpand\hyper@t@mp@=\relax
                                       \noexpand\hypertarget{\hyper@t@mp@}}%
                     \global\hypert@ks={\hyper@t@mp@}%
                     \else
                     \message\expandafter{\expandafter'\hyper@t@mp@' duplicate}%
		     \def\hyper@@tmp@{\hyperdef{#1}{#2}}%
		     \edef\hyper@@tmp@@{{#3@}}%
                     \global\let\hyper@t@mp@=\relax
                     \global\hypert@ks=\expandafter\expandafter
		     \expandafter{\expandafter\hyper@@tmp@\hyper@@tmp@@}%
                     \fi}\the\hypert@ks}%

\def\hyper@nique#1#2#3#4{{\escapechar=`\\\relax
                     \edef\hyper@t@mp@{\hyperstr@pquote#2.#3"\hyper@nd}%
                     \expandafter\ifx\csname hyperd@\meaning\hyper@t@mp@
                     \endcsname
                     \relax
                     \gdef#1{{}{\hyperstr@pquote#2"\hyper@nd}%
                               {\hyperstr@pquote#3"\hyper@nd}}%
                     \global\let\hyper@t@mp@=\relax
                     #4%
                     \else
		     \def\hyper@@tmp@{\hyper@nique{#1}{#2}}%
		     \edef\hyper@@tmp@@{{#3@}}%
                     \global\let\hyper@t@mp@=\relax
   		     \expandafter\hyper@@tmp@\hyper@@tmp@@{#4}%
                     \fi}}%

\ifx\protect\hyper@ndefined\let\protect=\relax\fi
\let\hyper@@@@=\relax
\def\hyper@@{\let\hyper@@@=\relax}%
\hyper@@
\def\hyper@{\relax\let\hyper@@@\noexpand\hyper@\noexpand}%
\def\hyperpr@ref{\hyper@@\hyperref}
\def\hyperpr@link{\hyper@@\hyperlink}
\def\hyperpr@def{\hyper@@\hyperdef}
\let\hyper@marker=\relax
\def\hyper@@tokenize#1\hyper@marker{#1}
\def\hyper@tokenize{\expandafter\endgroup\hyper@@tokenize}
\def\hyperpr@tokenize{\hyper@@\hyper@tokenize}

\edef\href{\begingroup\catcode`\string @=11
            \hyper@\hyperpr@tokenize\hyper@\hyperpr@link
            \hyper@marker}
\let\hname\hypertarget
\def\allowoncemore{\def\hyper@utoprocess{\let\hyper@utoprocess=\hyper@ndefined
                                         \hypers@fe }}%
%
%
\hypers@fe
 

\input amssym.tex 

\def\unredoffs{}
\tolerance=1000\hfuzz=2pt\def\fontflag{cm}
\catcode`\@=11 
\ifx\hyperdef\UNd@FiNeD\def\hyperdef#1#2#3#4{#4}\def\hyperref#1#2#3#4{#4}\def\href#1#2{#2}\fi
\magnification=1200\unredoffs\baselineskip=16pt plus 2pt minus 1pt
\def\Title#1#2{\nopagenumbers\rightline{#1}%
\vskip 1in\centerline{#2}\vskip .5in\pageno=0}
\def\Date#1{\vfill\leftline{#1}\tenpoint\supereject%
\footline={\hss\tenrm\hyperdef\hypernoname{page}\folio\folio\hss}}%

{\count255=\time\divide\count255 by 60 \xdef\hourmin{\number\count255}
 \multiply\count255 by-60\advance\count255 by\time
 \xdef\hourmin{\hourmin:\ifnum\count255<10 0\fi\the\count255}
}
\def\date{\number\day.\number\month.\number\year\ at \hourmin}

\def\draft{\message{ DRAFTMODE }\def\draftdate{{\rm Preliminary draft: \date}}%
\headline={\hfil\draftdate}\writelabels\baselineskip=20pt plus 2pt minus 2pt}
\def\nolabels{\def\wrlabeL##1{}\def\eqlabeL##1{}\def\reflabeL##1{}}
\def\writelabels{\def\wrlabeL##1{\leavevmode\vadjust{\rlap{\smash%
{\line{{\escapechar=` \hfill\rlap{\sevenrm\hskip.03in\string##1}}}}}}}%
\def\eqlabeL##1{{\escapechar-1\rlap{\sevenrm\hskip.05in\string##1}}}%
\def\reflabeL##1{\noexpand\llap{\noexpand\sevenrm\string\string\string##1}}}
\nolabels

\global\newcount\secno \global\secno=0
\global\newcount\meqno \global\meqno=1
\def\s@csym{}

\def\newsec#1\par{\global\advance\secno by1%
{\toks0{#1}\message{(\the\secno. \the\toks0)}}%
\global\subsecno=0\eqnres@t\let\s@csym\secsym\xdef\secn@m{\the\secno}\noindent
{\bf\hyperdef\hypernoname{section}{\the\secno}{\the\secno.} #1}%
\writetoca{{\string\hyperref{}{section}{\the\secno}{\bf \the\secno\quad}} {\bf #1}}\par%
\nobreak\medskip\nobreak\noindent\ignorespaces}
\def\eqnres@t{\xdef\secsym{\the\secno.}\global\meqno=1\bigbreak\bigskip}
\def\sequentialequations{\def\eqnres@t{\bigbreak}}\xdef\secsym{}

\global\newcount\subsecno \global\subsecno=0
\def\subsec#1\par{\global\advance\subsecno by1%
{\toks0{#1}\message{(\s@csym\the\subsecno. \the\toks0)}}%
\global\subsubsecno=0%
\ifnum\lastpenalty>9000\else\bigbreak\fi
\noindent{\it\hyperdef\hypernoname{subsection}{\secn@m.\the\subsecno}%
{\secn@m.\the\subsecno.} #1}\writetoca{\string\hskip1.45cm
{\string\hyperref{}{subsection}{\secn@m.\the\subsecno}{\secn@m.\the\subsecno.}}
{#1}}\par\nobreak\medskip\nobreak\noindent\ignorespaces}

\global\newcount\subsubsecno \global\subsubsecno=0
\def\subsubsec#1\par{\global\advance\subsubsecno by1%
{\toks0{#1}\message{(\secn@m.\the\subsecno.\the\subsubsecno. \the\toks0)}}%
\global\subsubsubsecno=0%
\ifnum\lastpenalty>9000\else\bigbreak\fi
\noindent{\it\hyperdef\hypernoname{subsubsection}{\secn@m.\the\subsecno\the\subsubsecno}%
{\secn@m.\the\subsecno.\the\subsubsecno.} #1}
\par\nobreak\medskip\nobreak\noindent\ignorespaces}

\global\newcount\subsubsubsecno \global\subsubsubsecno=0
\def\subsubsubsec#1\par{\global\advance\subsubsubsecno by1%
{\toks0{#1}\message{(\secn@m.\the\subsecno.\the\subsubsecno.\the\subsubsubsecno \the\toks0)}}%
\ifnum\lastpenalty>9000\else\bigbreak\fi
\noindent{\it\hyperdef\hypernoname{subsubsection}{\secn@m.\the\subsecno\the\subsubsecno\the\subsubsubsecno}%
{\secn@m.\the\subsecno.\the\subsubsecno.\the\subsubsubsecno.} #1}%
\par\nobreak\medskip\nobreak\noindent\ignorespaces}


\def\newnewsec#1#2\par{\global\advance\secno by1%
{\toks0{#2}\message{(\secn@m. \the\toks0)}}%
\global\subsecno=0\global\subsubsecno=0\eqnres@t\let\s@csym\secsym\xdef\secn@m{\the\secno}\noindent
\ifnum\lastpenalty>9000\else\bigbreak\fi
\noindent{\bf\hyperdef\hypernoname{section}{\secn@m}{\secn@m.} #2}%
\writetoca{{\string\hyperref{}{section}{\the\secno}{\bf \the\secno\quad}} {\bf #2}}
\DefWarn#1%
\xdef#1{\noexpand\hyperref{}{section}{\the\secno}%
{\the\secno}}\writedef{#1\leftbracket#1}\wrlabeL{#1=#1}%
\par\nobreak\medskip\nobreak\noindent\ignorespaces}

\def\newsubsec#1#2\par{\global\advance\subsecno by1%
{\toks0{#2}\message{(\secn@m.\the\subsecno. \the\toks0)}}%
\global\subsubsecno=0%
\ifnum\lastpenalty>9000\else\bigbreak\fi
\noindent{\it\hyperdef\hypernoname{subsection}{\secn@m.\the\subsecno}%
{\secn@m.\the\subsecno.} #2}
\DefWarn#1%
\xdef#1{\noexpand\hyperref{}{subsection}{\secn@m.\the\subsecno}%
{\secn@m.\the\subsecno}}\writedef{#1\leftbracket#1}\wrlabeL{#1=#1}%
\writetoca{\string\hskip1.45cm
{\string\hyperref{}{subsection}{\secn@m.\the\subsecno}{\secn@m.\the\subsecno.}}
{#2}}%
\par\nobreak\medskip\nobreak\noindent\ignorespaces}

\def\newsubsecstar#1#2\par{\global\advance\subsecno by1%
{\toks0{#2}\message{(\secn@m.\the\subsecno. \the\toks0)}}%
\global\subsubsecno=0%
\ifnum\lastpenalty>9000\else\bigbreak\fi
\noindent{\it\hyperdef\hypernoname{subsection}{\secn@m.\the\subsecno}%
{\secn@m.\the\subsecno.} #2}
\DefWarn#1%
\xdef#1{\noexpand\hyperref{}{subsection}{\secn@m.\the\subsecno}%
{\secn@m.\the\subsecno}}\writedef{#1\leftbracket#1}\wrlabeL{#1=#1}%
\par\nobreak\medskip\nobreak\noindent\ignorespaces}

\def\newsubsubsec#1#2\par{\global\advance\subsubsecno by1%
{\toks0{#2}\message{(\secn@m.\the\subsecno.\the\subsubsecno. \the\toks0)}}%
\global\subsubsubsecno=0%
\ifnum\lastpenalty>9000\else\bigbreak\fi
\noindent{\it\hyperdef\hypernoname{subsubsection}{\secn@m.\the\subsecno.\the\subsubsecno}%
{\secn@m.\the\subsecno.\the\subsubsecno.} #2}
\DefWarn#1%
\xdef#1{\noexpand\hyperref{}{subsubsection}{\secn@m.\the\subsecno.\the\subsubsecno}%
{\secn@m.\the\subsecno.\the\subsubsecno}}\writedef{#1\leftbracket#1}\wrlabeL{#1=#1}%
\par\nobreak\medskip\nobreak\noindent\ignorespaces}

\def\newsubsubsubsec#1#2\par{\global\advance\subsubsubsecno by1%
{\toks0{#2}\message{(\secn@m.\the\subsecno.\the\subsubsecno.\the\subsubsubsecno \the\toks0)}}%
\ifnum\lastpenalty>9000\else\bigbreak\fi
\noindent{\it\hyperdef\hypernoname{subsubsection}{\secn@m.\the\subsecno\the\subsubsecno\the\subsubsubsecno}%
{\secn@m.\the\subsecno.\the\subsubsecno.\the\subsubsubsecno.} #2}
\DefWarn#1%
\xdef#1{\noexpand\hyperref{}{subsubsubsection}{\secn@m.\the\subsecno.\the\subsubsecno.\the\subsubsubsecno}%
{\secn@m.\the\subsecno.\the\subsubsecno.\the\subsubsubsecno}}\writedef{#1\leftbracket#1}\wrlabeL{#1=#1}%
\par\nobreak\medskip\nobreak\noindent\ignorespaces}

\def\appendix#1#2{\global\meqno=1\global\subsecno=0\global\subsubsecno=0\xdef\secsym{\hbox{#1.}}%
\bigbreak\bigskip\noindent{\bf Appendix \hyperdef\hypernoname{appendix}{#1}%
{#1.} #2}{\toks0{(#1. #2)}\message{\the\toks0}}%
\xdef\s@csym{#1.}\xdef\secn@m{#1}%
\writetoca{{\string\hyperref{}{appendix}{#1}{\bf {#1}\quad}} {\bf #2}}%
\par\nobreak\medskip\nobreak}

%
\def\checkm@de#1#2{\ifmmode{\def\f@rst##1{##1}\hyperdef\hypernoname{equation}%
{#1}{#2}}\else\hyperref{}{equation}{#1}{#2}\fi}
\def\eqnn#1{\DefWarn#1\xdef #1{(\noexpand\relax\noexpand\checkm@de%
{\s@csym\the\meqno}{\secsym\the\meqno})}%
\wrlabeL#1\writedef{#1\leftbracket#1}\global\advance\meqno by1}
\def\f@rst#1{\c@t#1a\em@ark}\def\c@t#1#2\em@ark{#1}
\def\eqna#1{\DefWarn#1\wrlabeL{#1$\{\}$}%
\xdef #1##1{(\noexpand\relax\noexpand\checkm@de%
{\s@csym\the\meqno\noexpand\f@rst{##1}1}{\hbox{$\secsym\the\meqno##1$}})}
\writedef{#1\numbersign1\leftbracket#1{\numbersign1}}\global\advance\meqno by1}
\def\eqn#1#2{\DefWarn#1%
\xdef #1{(\noexpand\hyperref{}{equation}{\s@csym\the\meqno}%
{\secsym\the\meqno})}$$#2\eqno(\hyperdef\hypernoname{equation}%
{\s@csym\the\meqno}{\secsym\the\meqno})\eqlabeL#1$$%
\writedef{#1\leftbracket#1}\global\advance\meqno by1}
\def\xeqn{\expandafter\xe@n}\def\xe@n(#1){#1}
\def\xeqna#1{\expandafter\xe@n#1}
\def\eqns#1{(\e@ns #1{\hbox{}})}
\def\e@ns#1{\ifx\UNd@FiNeD#1\message{eqnlabel \string#1 is undefined.}%
\xdef#1{(?.?)}\fi{\let\hyperref=\relax\xdef\next{#1}}%
\ifx\next\em@rk\def\next{}\else%
\ifx\next#1\xeqn#1\else\def\n@xt{#1}\ifx\n@xt\next#1\else\xeqna#1\fi
\fi\let\next=\e@ns\fi\next}
\def\etag#1{\eqnn#1\eqno#1}\def\etaga#1{\eqna#1\eqno#1}
\def\DefWarn#1{}
%
\newskip\footskip\footskip14pt plus 1pt minus 1pt 
\def\footnotefont{\ninepoint}\def\f@t#1{\footnotefont #1\@foot}
\def\f@@t{\baselineskip\footskip\bgroup\footnotefont\aftergroup\@foot\let\next}
\setbox\strutbox=\hbox{\vrule height9.5pt depth4.5pt width0pt}
\global\newcount\ftno \global\ftno=0
\def\foot{\global\advance\ftno by1\def\foot@rg{\hyperref{}{footnote}%
{\the\ftno}{\the\ftno}\xdef\foot@rg{\noexpand\hyperdef\noexpand\hypernoname%
{footnote}{\the\ftno}{\the\ftno}}}\footnote{$^{\foot@rg}$}}
%
%
%
\global\newcount\refno \global\refno=1
\newwrite\rfile
\def\ref{[\hyperref{}{reference}{\the\refno}{\the\refno}]\nref}
\def\nref#1{\DefWarn#1%
\xdef#1{[\noexpand\hyperref{}{reference}{\the\refno}{\the\refno}]}%
\writedef{#1\leftbracket#1}%
\ifnum\refno=1\immediate\openout\rfile=\jobname.refs\fi
\chardef\wfile=\rfile\immediate\write\rfile{\noexpand\item{[\noexpand\hyperdef%
\noexpand\hypernoname{reference}{\the\refno}{\the\refno}]\ }%
\reflabeL{#1\hskip.31in}\pctsign}\global\advance\refno by1\findarg}
\def\findarg#1#{\begingroup\obeylines\newlinechar=`\^^M\pass@rg}
{\obeylines\gdef\pass@rg#1{\writ@line\relax #1^^M\hbox{}^^M}%
\gdef\writ@line#1^^M{\expandafter\toks0\expandafter{\striprel@x #1}%
\edef\next{\the\toks0}\ifx\next\em@rk\let\next=\endgroup\else\ifx\next\empty%
\else\immediate\write\wfile{\the\toks0}\fi\let\next=\writ@line\fi\next\relax}}
\def\striprel@x#1{} \def\em@rk{\hbox{}}
\def\lref{\begingroup\obeylines\lr@f}
\def\lr@f#1#2{\DefWarn#1\gdef#1{\let#1=\UNd@FiNeD\ref#1{#2}}\endgroup\unskip}
\def\semi{;\hfil\break}
\def\addref#1{\immediate\write\rfile{\noexpand\item{}#1}} 
\def\listrefs{\vfill\supereject\immediate\closeout\rfile\writestoppt
\baselineskip=\footskip\centerline{{\bf References}}\bigskip{\parindent=20pt%
\frenchspacing\escapechar=` \input \jobname.refs\vfill\eject}\nonfrenchspacing}
\def\startrefs#1{\immediate\openout\rfile=\jobname.refs\refno=#1}
\def\xref{\expandafter\xr@f}\def\xr@f[#1]{#1}
\def\refs#1{\count255=1[\r@fs #1{\hbox{}}]}
\def\r@fs#1{\ifx\UNd@FiNeD#1\message{reflabel \string#1 is undefined.}%
\nref#1{need to supply reference \string#1.}\fi%
\vphantom{\hphantom{#1}}{\let\hyperref=\relax\xdef\next{#1}}%
\ifx\next\em@rk\def\next{}%
\else\ifx\next#1\ifodd\count255\relax\xref#1\count255=0\fi%
\else#1\count255=1\fi\let\next=\r@fs\fi\next}
\def\figures{\centerline{{\bf Figure Captions}}\medskip\parindent=40pt%
\def\fig##1##2{\medskip\item{Fig.~\hyperdef\hypernoname{figure}{##1}{##1}.  }%
##2}}
%
\newwrite\ffile\global\newcount\figno \global\figno=1
\def\fig{fig.~\hyperref{}{figure}{\the\figno}{\the\figno}\nfig}
\def\nfig#1{\DefWarn#1%
\xdef#1{fig.~\noexpand\hyperref{}{figure}{\the\figno}{\the\figno}}%
\writedef{#1\leftbracket fig.\noexpand~\xfig#1}%
\ifnum\figno=1\immediate\openout\ffile=\jobname.figs\fi\chardef\wfile=\ffile%
{\let\hyperref=\relax
\immediate\write\ffile{\noexpand\medskip\noexpand\item{Fig.\ %
\noexpand\hyperdef\noexpand\hypernoname{figure}{\the\figno}{\the\figno}. }
\reflabeL{#1\hskip.55in}\pctsign}}\global\advance\figno by1\findarg}
\def\xfig{\expandafter\xf@g}\def\xf@g fig.\penalty\@M\ {}
\def\figs#1{figs.~\f@gs #1{\hbox{}}}
\def\f@gs#1{{\let\hyperref=\relax\xdef\next{#1}}\ifx\next\em@rk\def\next{}\else
\ifx\next#1\xfig #1\else#1\fi\let\next=\f@gs\fi\next}
%
\def\figin{\epsfcheck\figin}\def\figins{\epsfcheck\figins}
\def\epsfcheck{\ifx\epsfbox\UnDeFiNeD
\message{(NO epsf.tex, FIGURES WILL BE IGNORED)}
\gdef\figin##1{\vskip2in}\gdef\figins##1{\hskip.5in}
\else\message{(FIGURES WILL BE INCLUDED)}%
\gdef\figin##1{##1}\gdef\figins##1{##1}\fi}
\def\figinsert{\goodbreak\topinsert}
\def\ifig#1#2#3{\DefWarn#1\xdef#1{fig.~\the\figno}
\writedef{#1\leftbracket fig.\noexpand~\the\figno}%
\figinsert\figin{\centerline{#3}}
\smallskip
\leftskip=0pt \rightskip=0pt
\baselineskip12pt\noindent
{{\bf Fig.~\the\figno}\ \ninepoint #2}
\medskip
\global\advance\figno by1\par\endinsert}
\newwrite\lfile
{\escapechar-1\xdef\pctsign{\string\%}\xdef\leftbracket{\string\{}
\xdef\rightbracket{\string\}}\xdef\numbersign{\string\#}}
\def\writedefs{\immediate\openout\lfile=label.defs \def\writedef##1{%
{\let\hyperref=\relax\let\hyperdef=\relax\let\hypernoname=\relax
 \immediate\write\lfile{\string\checkdef\string##1\rightbracket}}}}%
\def\writestop{\def\writestoppt{\immediate\write\lfile{\string\pageno
 \the\pageno\string\startrefs\leftbracket\the\refno\rightbracket
 \string\def\string\secsym\leftbracket\secsym\rightbracket
 \string\secno\the\secno\string\meqno\the\meqno}\immediate\closeout\lfile}}
\def\writestoppt{}\def\writedef#1{}

\def\seclab#1\par{\DefWarn#1%
\xdef #1{\noexpand\hyperref{}{section}{\the\secno}{\the\secno}}%
\writedef{#1\leftbracket#1}\wrlabeL{#1=#1}\par%
\nobreak\medskip\nobreak\noindent\ignorespaces}
\def\subseclab#1\par{\DefWarn#1%
\xdef #1{\noexpand\hyperref{}{subsection}{\the\secno.\the\subsecno}%
{\the\secno.\the\subsecno}}\writedef{#1\leftbracket#1}\wrlabeL{#1=#1}\par%
\nobreak\medskip\nobreak\noindent\ignorespaces}
\def\subsubseclab#1\par{\DefWarn#1%
\xdef#1{\noexpand\hyperref{}{subsubsection}{\the\secno.\the\subsecno.\the\subsubsecno}%
{\the\secno.\the\subsecno.\the\subsubsecno}}\writedef{#1\leftbracket#1}\wrlabeL{#1=#1}\par%
\nobreak\medskip\nobreak\noindent\ignorespaces}
\def\applab#1\par{\DefWarn#1%
\xdef#1{\noexpand\hyperref{}{appendix}{\secn@m}{\secn@m}}%
\writedef{#1\leftbracket#1}\wrlabeL{#1=#1}%
\par\nobreak\medskip\nobreak\noindent\ignorespaces}
\def\appsublab#1{\DefWarn#1%
\xdef #1{\noexpand\hyperref{}{appendix}{\secn@m.\the\subsecno}{\secn@m.\the\subsecno}}%
\writedef{#1\leftbracket#1}\wrlabeL{#1=#1}}
\newwrite\tfile \def\writetoca#1{}
\def\leaderfill{\leaders\hbox to 1em{\hss.\hss}\hfill}
\def\writetoc{\immediate\openout\tfile=\jobname.toc
   \def\writetoca##1{{\edef\next{\write\tfile{\noindent ##1
   \string\leaderfill{
   \string\hyperref{}{page}{\noexpand\number\pageno}%
   {\noexpand\number\pageno}} \par}}\next}}
}
\newread\ch@ckfile
\def\listtoc{\immediate\closeout\tfile\immediate\openin\ch@ckfile=\jobname.toc
\ifeof\ch@ckfile\message{no file \jobname.toc, no table of contents this pass}%
\else\closein\ch@ckfile\centerline{\bf Contents}\nobreak\medskip%
{\baselineskip=15.5pt\footnotefont\parskip=0pt\catcode`\@=11\input\jobname.toc
\catcode`\@=12\bigbreak\bigskip}\fi}
\catcode`\@=12 
\font\authorfont=cmcsc10
\def\tenpoint{\def\rm{\fam0\tenrm}
\textfont0=\tenrm \scriptfont0=\sevenrm \scriptscriptfont0=\fiverm
\textfont1=\teni  \scriptfont1=\seveni  \scriptscriptfont1=\fivei
\textfont2=\tensy \scriptfont2=\sevensy \scriptscriptfont2=\fivesy
\textfont\itfam=\tenit \def\it{\fam\itfam\tenit}\def\footnotefont{\ninepoint}%
\textfont\bffam=\tenbf \def\bf{\fam\bffam\tenbf}\def\sl{\fam\slfam\tensl}\rm}
\font\ninerm=cmr9 \font\sixrm=cmr6 \font\ninei=cmmi9 \font\sixi=cmmi6
\font\ninesy=cmsy9 \font\sixsy=cmsy6 \font\ninebf=cmbx9
\font\nineit=cmti9 \font\ninesl=cmsl9 \skewchar\ninei='177
\skewchar\sixi='177 \skewchar\ninesy='60 \skewchar\sixsy='60
\def\ninepoint{\def\rm{\fam0\ninerm}
\textfont0=\ninerm \scriptfont0=\sixrm \scriptscriptfont0=\fiverm
\textfont1=\ninei \scriptfont1=\sixi \scriptscriptfont1=\fivei
\textfont2=\ninesy \scriptfont2=\sixsy \scriptscriptfont2=\fivesy
\textfont\itfam=\ninei \def\it{\fam\itfam\nineit}\def\sl{\fam\slfam\ninesl}%
\textfont\bffam=\ninebf \def\bf{\fam\bffam\ninebf}\rm}
%
\hyphenation{anom-aly anom-alies coun-ter-term coun-ter-terms}

\def\tikzcaption#1#2{\DefWarn#1\xdef#1{Fig.~\the\figno}
\writedef{#1\leftbracket Fig.\noexpand~\the\figno}%
{
\smallskip
\leftskip=20pt \rightskip=20pt \baselineskip12pt\noindent
{{\bf Fig.~\the\figno}\ \ninepoint #2}
\bigskip
\global\advance\figno by1 \par}}

\def\ntoalpha#1{%
\ifcase#1%
@%
\or A\or B\or C\or D\or E\or F\or G\or H\or I\or J\or K\or L\or M%
\fi
}

\global\newcount\appno \global\appno=1
\def\applab#1{\xdef #1{\ntoalpha{\appno}}\writedef{#1\leftbracket#1}\wrlabeL{#1=#1}
\global\advance\appno by1}

\def\preprint#1 #2\par{\rightline{\vbox{\baselineskip12pt\hbox{#1}\hbox{#2}}}\vskip2cm}
%
\def\title#1\par{\centerline{\bf #1}\nopagenumbers\pageno=0}
\def\author#1\par{\bigskip\bigskip\centerline{#1}}

\newcount\addressno

\def\email#1#2{
\footnote{\null}{\kern-\parindent \llap{$^#1$\hskip1pt}email: #2}}

\def\startcenter{%
  \par
  \begingroup
  \leftskip=0pt plus 1fil
  \rightskip=\leftskip
  \parindent=0pt
  \parfillskip=0pt
}
\def\stopcenter{\endgroup}

\def\address{\bigskip%
  \ifnum\the\addressno=0\else\stopcenter\endgroup\fi
  \advance\addressno by 1%
  \begingroup
  \startcenter
  \it
  \obeylines
  \addressAux
}
\def\addressAux#1{#1}

\def\abstract{\stopcenter\endgroup\bigskip\bigskip\noindent}

\def\Dsl{\,\raise.15ex\hbox{/}\mkern-13.5mu D} 
\def\dsl{\raise.15ex\hbox{/}\kern-.57em\partial}
\def\tr{{\rm tr}} \def\Tr{{\rm Tr}}
\def\boxeqn#1{\vcenter{\vbox{\hrule\hbox{\vrule\kern3pt\vbox{\kern3pt
	\hbox{${\displaystyle #1}$}\kern3pt}\kern3pt\vrule}\hrule}}}
\def\grad#1{\,\nabla\!_{{#1}}\,}
\def\gradgrad#1#2{\,\nabla\!_{{#1}}\nabla\!_{{#2}}\,}
\def\lform{\hbox{$\sqcup$}\llap{\hbox{$\sqcap$}}}
\def\lie{\hbox{\it\$}} 

\def\ap{{\alpha^{\prime}}}
\def\halfap#1{\Big({\ap\over 2}\Big)^{\mkern-4mu #1}}
\def\a{\alpha}
\def\b{{\beta}}
\def\g{{\gamma}}
\def\d{{\delta}}
\def\e{{\epsilon}}
\def\l{\lambda}
\def\k{{\kappa}}
\def\s{{\sigma}}
\def\t{{\theta}}
\def\om{{\omega}}
\def\lb{{\overline\lambda}}
\def\llb{(\l\lb)}
\def\wb{{\overline w}}
\def\half{{1\over 2}}
\def\p{{\partial}}
\def\pb{{\overline\partial}}
\def\tb{{\overline\theta}}
\def\bar{\overline}
\def\({\left(}
\def\){\right)}
\def\dz{{\rm d}z}

\def\cA{{\cal A}}
\def\cF{{\cal F}}
\def\cJ{{\cal J}}
\def\cK{{\cal K}}
\def\cI{{\cal I}}
\def\cV{{\cal V}}
\def\cW{{\cal W}}
\def\cY{{\cal Y}}
\def\cZ{{\cal Z}}

\def\bV{{\Bbb V}}
\def\bA{{\Bbb A}}
\def\bW{{\Bbb W}}
\def\bF{{\Bbb F}}

\def\Box{\square}
\def\AYM{A^{\rm SYM}}

\def\psum{\mathop{\sum\nolimits'}}

\def\len#1{{%
\def\Dlen{\left|\mkern-1mu #1\mkern -0.5mu\right|}%
\def\Sslen{\left|\mkern-1.3mu #1\mkern -1.3mu\right|}%
\def\SSlen{\left|\mkern-2.8mu #1\mkern-1.3mu\right|}%
\mathchoice{\Dlen}{\Dlen}{\Sslen}{\SSlen}}}

\def\perm#1{{\rm perm}#1}
\def\eikx{{\bigl\langle \prod_{j=1}^4 {\rm e}^{i k^j\cdot x^j}\bigr\rangle}}
\def\ImOmega{\Im\Omega}
\def\lalb#1{(\l\g^{#1}\lb)}
\def\Im{\mathop{{\rm Im}}} 
\def\sfrac#1/#2{\kern.1em\raise.5ex\hbox{\the\scriptfont0 #1}%
\kern-.1em/\kern-.15em\lower.25ex\hbox{\the\scriptfont0 #2}}

\font\tenshuffle=shuffle10 \font\sevenshuffle=shuffle7 \font\fiveshuffle=shuffle7 at 5pt
\def\shuffle{{%
\def\Dshuffle{\mathbin{\hbox{\tenshuffle\char'001}}}%
\def\Sshuffle{\mathbin{\hbox{\sevenshuffle\char'001}}}%
\def\SSshuffle{\mathbin{\hbox{\fiveshuffle\char'001}}}%
\mathchoice{\Dshuffle}{\Dshuffle}{\Sshuffle}{\SSshuffle}}}

\font\tenwedge=stmary10 \font\sevenwedge=stmary7 \font\fivewedge=stmary5
\def\owedge{%
\def\Dowedge{\mathbin{\hbox{\tenwedge\char"3F}}}%
\def\Sowedge{\mathbin{\hbox{\sevenwedge\char"3F}}}%
\def\SSowedge{\mathbin{\hbox{\fivewedge\char"3F}}}%
\mathchoice{\Dowedge}{\Dowedge}{\Sowedge}{\SSowedge}}

\def\qed{\hbox{\hskip 3pt
\vbox{\hrule\hbox to 7pt{\vrule height 7pt\hfill\vrule}
\hrule}}\hskip3pt}

\overfullrule=0pt\relax

\frenchspacing

\def\checkdef#1#2{
\ifx\UndeFined#1%
	\def#1{#2}
\else
	\immediate\write16{*** BUG ***: the label \string#1 is already defined ***}
\fi
}
\newread\instream
\def\openlabels{
\openin\instream= label.defs
\ifeof\instream\message{No labels in advance yet. Wait till next pass.}
\else\closein\instream \input label.defs
\fi}

\openin\instream= label.defs
\ifeof\instream\message{No labels in advance yet. Wait till next pass.}
\else\closein\instream \input label.defs
\fi
\writedefs

\def\arXiv:#1].{\hepthStrip#1 \nil}
\def\hepthStrip#1 #2\nil{\href{http://arxiv.org/abs/#1}{arXiv:#1 #2\unskip}].}

\font\frakfont=eufm10 at 10pt
\def\ce{\mathord{\hbox{\frakfont e}}}
\def\cf{\mathord{\hbox{\frakfont f}}}
\def\cm{\mathord{\hbox{\frakfont M}}}

\input amssym

\catcode`\@11\relax
\def\youngdim{1.6}
\newif\ify@autoscale \y@autoscaletrue \def\Yautoscale#1{\ifnum #1=0
  \y@autoscalefalse\else\y@autoscaletrue\fi}
\newdimen\y@b@xdim
\newdimen\y@boxdim \y@boxdim=13pt
\def\Yboxdim#1{\y@autoscalefalse\y@boxdim=#1}
\newdimen\y@linethick    \y@linethick=.3pt
\def\Ylinethick#1{\y@linethick=#1}
\newskip\y@interspace \y@interspace=0ex plus 0.3ex
\def\Yinterspace#1{\y@interspace=#1}
\newif\ify@vcenter   \y@vcenterfalse
\def\Yvcentermath#1{\ifnum #1=0 \y@vcenterfalse\else\y@vcentertrue\fi}
\newif\ify@stdtext   \y@stdtextfalse
\def\Ystdtext#1{\ifnum #1=0 \y@stdtextfalse\else\y@stdtexttrue\fi}
\newif\ify@enable@skew   \y@enable@skewfalse
\expandafter\ifx\csname enableskew\endcsname\relax
 \y@enable@skewfalse \else \y@enable@skewtrue\fi
\def\y@vr{\vrule height0.8\y@b@xdim width\y@linethick depth 0.2\y@b@xdim}
\def\y@emptybox{\y@vr\hbox to \y@b@xdim{\hfil}}
\ify@enable@skew
 \def\y@abcbox#1{\if :#1\else
   \y@vr\hbox to \y@b@xdim{\hfil#1\hfil}\fi}
 \def\y@mathabcbox#1{\if :#1\else
   \y@vr\hbox to \y@b@xdim{\hfil$#1$\hfil}\fi}
\else
 \def\y@abcbox#1{\y@vr\hbox to \y@b@xdim{\hfil#1\hfil}}
 \def\y@mathabcbox#1{\y@vr\hbox to \y@b@xdim{\hfil$#1$\hfil}}
\fi
\def\y@setdim{%
  \ify@autoscale%
   \ifvoid1\else\typeout{Package youngtab: box1 not free! Expect an
     error!}\fi%
   \setbox1=\hbox{A}\y@b@xdim=\youngdim\ht1 \setbox1=\hbox{}\box1%
  \else\y@b@xdim=\y@boxdim \advance\y@b@xdim by -2\y@linethick
  \fi}
\newcount\y@counter
\newif\ify@islastarg
\def\y@lastargtest#1,#2 {\if\space #2 \y@islastargtrue
  \else\y@islastargfalse\fi}
\def\y@emptyboxes#1{\y@counter=#1\loop\ifnum\y@counter>0
  \advance\y@counter by -1 \y@emptybox\repeat}
\def\y@nelineemptyboxes#1{%
  \vbox{%
    \hrule height\y@linethick%
    \hbox{\y@emptyboxes{#1}\y@vr}
    \hrule height\y@linethick}\vskip-\y@linethick}
\def\yng(#1){%
  \y@setdim%
  \hskip\y@interspace%
  \ifmmode\ify@vcenter\vcenter\fi\fi{%
  \y@lastargtest#1,
  \vbox{\offinterlineskip
    \ify@islastarg
     \y@nelineemptyboxes{#1}
    \else
     \y@ungempty(#1)
    \fi}}\hskip\y@interspace}
\def\y@ungempty(#1,#2){%
  \y@nelineemptyboxes{#1}
  \y@lastargtest#2,
  \ify@islastarg
   \y@nelineemptyboxes{#2}
  \else
   \y@ungempty(#2)
  \fi}
\def\y@nelettertest#1#2. {\if\space #2 \y@islastargtrue
  \else\y@islastargfalse\fi}
\def\y@abcboxes#1#2.{%
  \ify@stdtext\y@abcbox#1\else\y@mathabcbox#1\fi%
  \y@nelettertest #2.
  \ify@islastarg\unskip%
   \ify@stdtext\y@abcbox{#2}\else\y@mathabcbox{#2}\fi%
  \else\y@abcboxes#2.\fi}
 \newdimen\y@full@b@xdim
 \newcount\y@m@veright@cnt
\ify@enable@skew
 \def\y@get@m@veright@cnt#1#2.{%
   \if :#1 \advance\y@m@veright@cnt by 1\y@get@m@veright@cnt#2.\fi}
 \let\y@setdim@=\y@setdim
 \def\y@setdim{%
   \y@setdim@ \y@full@b@xdim=\y@b@xdim
   \advance\y@full@b@xdim by 1\y@linethick}
 \def\y@m@veright@ifskew#1{
   \y@m@veright@cnt=0 \y@get@m@veright@cnt#1.
   \moveright \y@m@veright@cnt\y@full@b@xdim}
\else
 \def\y@m@veright@ifskew#1{}
\fi
\def\y@nelineabcboxes#1{%
  \y@nelettertest #1.
  \ify@islastarg
   \y@m@veright@ifskew{#1}
    \vbox{
      \hrule height\y@linethick%
      \hbox{\ify@stdtext\y@abcbox#1\else\y@mathabcbox#1\fi\y@vr}
      \hrule height\y@linethick}\vskip-\y@linethick
  \else
   \y@m@veright@ifskew{#1}
    \vbox{
      \hrule height\y@linethick%
      \hbox{\y@abcboxes #1.\y@vr}%
      \hrule height\y@linethick}\vskip-\y@linethick
  \fi}
\def\young(#1){%
  \y@setdim%
  \hskip\y@interspace%
  \y@lastargtest#1,
  \ifmmode\ify@vcenter\vcenter\fi\fi{%
  \vbox{\offinterlineskip
    \ify@islastarg\y@nelineabcboxes{#1}%
    \else\y@ungabc(#1)%
    \fi}}\hskip\y@interspace}
\def\y@ungabc(#1,#2){%
  \y@nelineabcboxes{#1}%
  \y@lastargtest#2,
  \ify@islastarg\y@nelineabcboxes{#2}%
  \else\y@ungabc(#2)%
  \fi}
\catcode`\@12\relax
 

\def\textbf#1{{\bf #1}}
\def\paragraph#1{\medskip\noindent{\it #1.}}
\def\cX{{\cal X}}
\def\u#1{{\underline #1}}
\def\cyc{{\rm cyclic}}


\title The massive one-loop four-point string amplitude

\title in pure spinor superspace

\author
Carlos R. Mafra\email{{}}{c.r.mafra@soton.ac.uk}

\address
Mathematical Sciences and STAG Research Centre, University of Southampton,
Highfield, Southampton, SO17 1BJ, UK

\abstract
The open- and closed-string three- and four-point one-loop amplitudes involving massless states
and one first-level massive state
are
computed in pure spinor superspace. For the open string, we show that their one-loop correlators
can be rewritten in terms of tree-level kinematic factors.
We then analyze the closed string. For three points, this is immediate.
For four points, we show that it is possible to rewrite the one-loop closed-string correlator
using tree-level kinematic factors, but only for
certain combinations of massive and massless
states (different for type IIA and IIB).

\Date{August 2025}

\lref\UV{
	S.~P.~Kashyap, C.~R.~Mafra, M.~Verma and L.~Ypanaqu{\'e},
	``Massless representation of massive superfields and tree amplitudes with the pure spinor formalism,''
	JHEP \textbf{02}, 215 (2025)
	[\arXiv:2407.02436 [hep-th]].
}

\lref\HarnadS{
	J.P.~Harnad and S.~Shnider,
  	``Constraints And Field Equations For Ten-dimensional Superyang-mills Theory,''
	Commun.\ Math.\ Phys.\  {\bf 106}, 183 (1986).
}

\lref\adynkras{
	S.~J.~Gates, Y.~Hu and S.~N.~H.~Mak,
	``Advening to adynkrafields: Young tableaux to component fields of the 10D, ${\cal N}=1$ scalar superfield,''
	Adv. Theor. Math. Phys. \textbf{25}, no.6, 1449-1547 (2021)
	[\arXiv:2006.03609 [hep-th]].
}

\lref\wip{
	C. Huang, C.R. Mafra, Yi-Xiao Tao, ``work in progress''
}

\lref\sagan{
	B.E.~Sagan, ``The symmetric group; representations, combinatorial algorithms and symmetric
	functions'' 2nd edition, Springer, 2001.
}
\lref\gdjames{
	G.D.~James, ``The Representation Theory of the Symmetric Groups'', Springer, 1978.
}

\lref\LieArt{
	R.~Feger, T.~W.~Kephart and R.~J.~Saskowski,
	``LieART 2.0 --- A Mathematica application for Lie Algebras and Representation Theory,''
	Comput. Phys. Commun. \textbf{257}, 107490 (2020)
	[\arXiv:1912.10969 [hep-th]].
}

\lref\bianchihet{
	M.~Bianchi, L.~Lopez and R.~Richter,
	``On stable higher spin states in Heterotic String Theories,''
	JHEP \textbf{03}, 051 (2011)
	[\arXiv:1010.1177 [hep-th]].
}

\lref\FultonYoungbook{
	W.~Fulton, ``Young tableaux'', vol. 35 of London Mathematical Society Student Texts. Cambridge
	University Press, Cambridge, 1997. With applications to representation theory and geometry.
}

\lref\fischler{
	M.~Fischler,
	``Young-tableau methods for Kronecker products of representations of the classical groups,''
	J. Math. Phys. 22, 637 (1981)
}

\lref\robbins{
	F.~Rejon-Barrera and D.~Robbins,
	``Scalar-Vector Bootstrap,''
	JHEP \textbf{01}, 139 (2016)
	[\arXiv:1508.02676 [hep-th]].
}

\lref\hansenI{
M.~S.~Costa, T.~Hansen, J.~Penedones and E.~Trevisani,
``Projectors and seed conformal blocks for traceless mixed-symmetry tensors,''
JHEP \textbf{07}, 018 (2016)
[\arXiv:1603.05551 [hep-th]].
}
\lref\hansenII{
M.~S.~Costa and T.~Hansen,
``Conformal correlators of mixed-symmetry tensors,''
JHEP \textbf{02}, 151 (2015)
[\arXiv:1411.7351 [hep-th]].
}
\lref\hansenIII{
A.~Antunes, M.~S.~Costa, T.~Hansen, A.~Salgarkar and S.~Sarkar,
``The perturbative CFT optical theorem and high-energy string scattering in AdS at one loop,''
JHEP \textbf{04}, 088 (2021)
[\arXiv:2012.01515 [hep-th]].
}

\lref\minahan{
	T.~Bargheer, J.~A.~Minahan and R.~Pereira,
	``Computing Three-Point Functions for Short Operators,''
	JHEP \textbf{03}, 096 (2014)
	[\arXiv:1311.7461 [hep-th]].\semi
	J.~A.~Minahan,
	``Holographic three-point functions for short operators,''
	JHEP \textbf{07}, 187 (2012)
	[\arXiv:1206.3129 [hep-th]].
}
\lref\stahngarnir{
	M.~B.~Green, K.~Peeters and C.~Stahn,
	``Superfield integrals in high dimensions,''
	JHEP \textbf{08}, 093 (2005)
	[\arXiv:hep-th/0506161 [hep-th]].
}

\lref\oneloopMichael{
	M.~B.~Green, C.~R.~Mafra and O.~Schlotterer,
  	``Multiparticle one-loop amplitudes and S-duality in closed superstring theory,''
	JHEP {\bf 1310}, 188 (2013).
	[\arXiv:1307.3534 [hep-th]].
}

\lref\wildenauer{
	B. Wildenauer, ``Superstring amplitudes at 1-loop including one
	massive state,’’ master thesis, Uppsala University (2024),\semi
	https://uu.diva-portal.org/smash/get/diva2:1920951/FULLTEXT01.pdf
}

\lref\GreenMN{
	M.~B.~Green, J.~H.~Schwarz and E.~Witten,
	``Superstring Theory. Vol. 2: Loop Amplitudes, Anomalies And Phenomenology,''
	Cambridge  University Press (1987).
}
\lref\verlindes{
	E.P.~Verlinde and H.L.~Verlinde,
	``Chiral Bosonization, Determinants and the String Partition Function,''
	Nucl.\ Phys.\ B {\bf 288}, 357 (1987).
}
\lref\LiEprogram{
	M.A.A. van Leeuwen, A.M. Cohen and B. Lisser,
	``LiE, A Package for Lie Group Computations'', Computer Algebra Nederland, Amsterdam, ISBN 90-74116-02-7, 1992
	\semi
	{\tt http://wwwmathlabo.univ-poitiers.fr/\~{}maavl/LiE/}
}
\lref\DHokerPDL{
	E.~D'Hoker and D.~H.~Phong,
  	``The Geometry of String Perturbation Theory,''
	Rev.\ Mod.\ Phys.\  {\bf 60}, 917 (1988).
}

\lref\xerox{
	E.~D'Hoker and D.~H.~Phong,
  	``Conformal Scalar Fields and Chiral Splitting on Superriemann Surfaces,''
	Commun.\ Math.\ Phys.\  {\bf 125}, 469 (1989).
}

\lref\massivewebsite{
	http://www.southampton.ac.uk/\~{}crm1n16/massive.html
}

\lref\ellipticMZV{
	J.~Broedel, C.R.~Mafra, N.~Matthes and O.~Schlotterer,
	``Elliptic multiple zeta values and one-loop superstring amplitudes,''
	JHEP \textbf{07}, 112 (2015)
	[\arXiv:1412.5535 [hep-th]].
}

\lref\nptMethod{
	C.~R.~Mafra, O.~Schlotterer, S.~Stieberger and D.~Tsimpis,
	``A recursive method for SYM n-point tree amplitudes,''
	Phys.\ Rev.\ D {\bf 83}, 126012 (2011).
	[\arXiv:1012.3981 [hep-th]].
}
\lref\towards{
	C.R.~Mafra,
	``Towards Field Theory Amplitudes From the Cohomology of Pure Spinor Superspace,''
	JHEP {\bf 1011}, 096 (2010).
	[\arXiv:1007.3639 [hep-th]].
}

\lref\website{
	http://www.southampton.ac.uk/\~{}crm1n16/pss.html
}
\lref\EOMbbs{
	C.R.~Mafra and O.~Schlotterer,
  	``Multiparticle SYM equations of motion and pure spinor BRST blocks,''
	JHEP {\bf 1407}, 153 (2014).
	[\arXiv:1404.4986 [hep-th]].
}

\lref\partIcohomology{
	C.R.~Mafra and O.~Schlotterer,
	``Cohomology foundations of one-loop amplitudes in pure spinor superspace,''
	[\arXiv:1408.3605 [hep-th]].
}

\lref\massSweden{
       	M.~Guillen, H.~Johansson, R.~L.~Jusinskas and O.~Schlotterer,
	``Scattering Massive String Resonances through Field-Theory Methods,''
	Phys. Rev. Lett. \textbf{127}, no.5, 051601 (2021)
	[\arXiv:2104.03314 [hep-th]].
}
\lref\oneloopbb{
	C.R.~Mafra and O.~Schlotterer,
	``The Structure of n-Point One-Loop Open Superstring Amplitudes,''
	JHEP \textbf{08}, 099 (2014)
	[\arXiv:1203.6215 [hep-th]].
}

\lref\MSSI{
	C.R.~Mafra, O.~Schlotterer and S.~Stieberger,
	``Complete N-Point Superstring Disk Amplitude I. Pure Spinor Computation,''
	Nucl.\ Phys.\ B {\bf 873}, 419 (2013).
	[\arXiv:1106.2645 [hep-th]].
}
\lref\MSSII{
	C.~R.~Mafra, O.~Schlotterer and S.~Stieberger,
	``Complete N-Point Superstring Disk Amplitude II. Amplitude
	and Hypergeometric Function Structure,''
	Nucl.\ Phys.\ B {\bf 873}, 461 (2013).
	[\arXiv:1106.2646 [hep-th]].
}
\lref\fourtree{
	C.R.~Mafra,
	``Pure Spinor Superspace Identities for Massless Four-point Kinematic Factors,''
	JHEP \textbf{04}, 093 (2008)
	[\arXiv:0801.0580 [hep-th]].
}
\lref\PSthreemass{
	S.~Chakrabarti, S.~P.~Kashyap and M.~Verma,
	``Amplitudes Involving Massive States Using Pure Spinor Formalism,''
	JHEP \textbf{12}, 071 (2018)
	[\arXiv:1808.08735 [hep-th]].
}
\lref\mafraids{
	C.R.~Mafra,
  	``Pure Spinor Superspace Identities for Massless Four-point Kinematic Factors,''
	JHEP {\bf 0804}, 093 (2008).
	[\arXiv:0801.0580 [hep-th]].
}
\lref\masstheta{
	S.~Chakrabarti, S.~P.~Kashyap and M.~Verma,
	``Theta Expansion of First Massive Vertex Operator in Pure Spinor,''
	JHEP \textbf{01}, 019 (2018)
	[\arXiv:1706.01196 [hep-th]].
}

\lref\PSS{
	C.R.~Mafra,
	``PSS: A FORM Program to Evaluate Pure Spinor Superspace Expressions,''
	[\arXiv:1007.4999 [hep-th]].
}

\lref\ICTP{
	N.~Berkovits,
  	``ICTP lectures on covariant quantization of the superstring,''
	[hep-th/0209059].
}

\lref\BCpaper{
	N.~Berkovits and O.~Chandia,
	``Massive superstring vertex operator in D = 10 superspace,''
	JHEP \textbf{08}, 040 (2002)
	[\arXiv:hep-th/0204121 [hep-th]].
}

\lref\psf{
 	N.~Berkovits,
	``Super-Poincare covariant quantization of the superstring,''
	JHEP {\bf 0004}, 018 (2000)
	[\arXiv:hep-th/0001035 [hep-th]].
}

\lref\wittentwistor{
	E.Witten,
        ``Twistor-Like Transform In Ten-Dimensions''
        Nucl.Phys. B {\bf 266}, 245~(1986)
}
\lref\higherSYM{
	C.R.~Mafra and O.~Schlotterer,
	``A solution to the non-linear equations of D=10 super Yang-Mills theory,''
	Phys.\ Rev.\ D {\bf 92}, no. 6, 066001 (2015).
	[\arXiv:1501.05562 [hep-th]].
}
\lref\treereview{
	C.~R.~Mafra and O.~Schlotterer,
	``Tree-level amplitudes from the pure spinor superstring,''
	Phys. Rept. \textbf{1020}, 1-162 (2023)
	[\arXiv:2210.14241 [hep-th]].
}

\lref\FORM{
	J.A.M.~Vermaseren,
	``New features of FORM,''
	arXiv:math-ph/0010025.
\semi
	M.~Tentyukov and J.A.M.~Vermaseren,
	``The multithreaded version of FORM,''
	arXiv:hep-ph/0702279.
}
\lref\MPS{
	N.~Berkovits,
  	``Multiloop amplitudes and vanishing theorems using the pure spinor formalism for the superstring,''
	JHEP {\bf 0409}, 047 (2004).
	[hep-th/0406055].
}

\lref\PSspace{
	N.~Berkovits,
	``Explaining Pure Spinor Superspace,''
	[\arXiv:hep-th/0612021 [hep-th]].
}

\lref\oneloopI{
	C.R.~Mafra and O.~Schlotterer,
	``Towards the n-point one-loop superstring amplitude. Part I. Pure spinors and superfield kinematics,''
	JHEP \textbf{08}, 090 (2019)
	[\arXiv:1812.10969 [hep-th]].
}
\lref\oneloopII{
	C.R.~Mafra and O.~Schlotterer,
	``Towards the n-point one-loop superstring amplitude. Part II. Worldsheet functions and their duality to kinematics,''
	JHEP \textbf{08}, 091 (2019)
	[\arXiv:1812.10970 [hep-th]].
}
\lref\oneloopIII{
	C.R.~Mafra and O.~Schlotterer,
	``Towards the n-point one-loop superstring amplitude. Part III. One-loop correlators and their double-copy structure,''
	JHEP \textbf{08}, 092 (2019)
	[\arXiv:1812.10971 [hep-th]].
}
\lref\towmass{
	C.R.~Mafra,
	``Towards massive field-theory amplitudes from the cohomology of pure spinor superspace,''
	JHEP \textbf{11}, 045 (2024)
	[\arXiv:2407.11849 [hep-th]].
}

\listtoc
\writetoc

\newsec{Introduction}

The main goal of this paper is to calculate the open-string one-loop correlators
of the three- and four-point string amplitudes with a single massive state of $({\rm
mass})^2 = 1/\ap$
using the pure spinor formalism \refs{\psf,\MPS}.
In order to do this, we will employ the same BRST-cohomology techniques
that were used in \refs{\oneloopI,\oneloopII,\oneloopIII}
to obtain the string
one-loop amplitudes up to seven massless external legs.

The answers
in pure spinor superspace \PSspace\ can be expressed in different ways, depending on which feature
is emphasized among BRST invariance, single-valuedness and locality. For instance, the three- and
four-point open-string massive correlators at one loop
\eqnn\Kcors
$$\eqalignno{
\cK_3 &= 
C_{\u1|2,3}\,,&\Kcors\cr
\cK_4 &=
s_{23}f^{(1)}_{23}C_{\u1|23,4}
+ s_{24}f^{(1)}_{24}C_{\u1|24,3}
+ s_{34}f^{(1)}_{34}C_{\u1|34,2}\,,
}$$
manifest BRST invariance and single-valuedness, where the massive state is
labelled by $\u1$ while the other labels $2,3,4$ represent the
massless
super-Yang-Mills states. In the above,
$C_{\u1| \ldots}$ are massive BRST invariants
defined below and $f^{(1)}_{ij}$ denote single-valued worldsheet functions on a genus-one surface
\refs{\ellipticMZV,\oneloopII}.

A BRST cohomology proof that the three-point massive one-loop {\it open-string} correlator is proportional to the
open-string kinematic factor $\cK^{\rm tree}_{\u1|2,3}$ at {\it tree level} is presented in the Appendix~\proofap.
Using the explicit polarizations and momenta extracted from its pure spinor superspace
expression\foot{Available to download in \massivewebsite.}
\refs{\PSS,\FORM},
the four-point {\it open-string} correlator at one loop is also shown to be rewritten in terms of
open-string kinematic factor $\cK^{\rm tree}_{\u1|2,3,4}$ at tree level, for the full supermultiplet; this
extends the earlier RNS analysis of \wildenauer.

We also investigate whether the {\it closed string} massive one-loop correlators can be
rewritten in terms of {\it tree-level} kinematic factors. For three points, this can be
clearly done. For four points, a vectorial contraction between left- and right-movers
present at one-loop and absent at tree-level has the potential to prevent this rewriting.
Rather surprisingly, it turns out that for certain combinations of massive and massless states
the one-loop correlator can be rewritten in terms of its tree-level counterpart. The details are
in section~\fptcorr. Finally, Appendix~\decapp\ reviews the mapping between $SO(n)$ Dynkin
labels and Young diagrams \fischler, their tensorial description as well as their symmetries.

\paragraph{Conventions} We use $s_{ij}=(k_i\cdot k_j)$ as a shorthand; these
are not proportional to the usual Mandelstam
variables when either $i$ or $j$ represents a massive leg. In addition, the pure spinor
bracket $\langle . \rangle$ \psf\ that extracts the component expansions
from the pure spinor superspace expressions is frequently omitted throughout.
Its presence is based on context; sometimes we
emphasize the BRST variations of expressions, sometimes we extract their components.
And finally, the word {\it massive} in this paper refers to a single massive state of the first
mass level.

\newnewsec\revsec Review

In this section we will briefly review the discussion of
section 2.1 in \oneloopI, and refer the reader to it for any missing details.

\newsubsec\pssec The pure spinor amplitude prescription

The one-loop amplitude prescription in the minimal pure spinor formalism is \MPS,
\eqn\onepresc{
{\cal A}_n =
\sum_{\rm top} C_{\rm top} \int_{D_{\rm top}}\!\!\!\!
d\tau \,\langle (\mu, b)\,\cZ\,
V_1(z_1)\prod_{j=2}^n \int\, dz_j\, U_j(z_j)\rangle\,,
}
where the Beltrami differential $\mu$ and the modulus $\tau$ encode the
topological information of the genus-one surface. The sum over the different genus-one
topologies (planar cylinder, M\"obius strip, and non-planar cylinder) implies
different integration domains $D_{\rm top}$ and color factors $C_{\rm top}$ for each, but in this
paper we will not focus on this aspect (see \GreenMN\ for details). Rather, we concentrate on the
CFT aspect of evaluating the correlation function of the picture-changing operators $\cZ$,
the $b$ ghost, and the vertex operators. More specifically, after integrating out all the non-zero modes
as well as the zero modes in \onepresc, we obtain an expression of the form
\eqn\theamp{
{\cal A}_n  =
\sum_{\rm top} C_{\rm top} \int_{D_{\rm top}}\!\!\!\!
d\tau \, dz_2 \, d z_3 \, \ldots \, d z_{n} \, \int d^{D} \ell \ |{\cal I}_n(\ell)| \,
\langle {\cal K}_n(\ell)  \rangle \, ,
}
where $\cK_n(\ell)$ is called the {\it correlator}, and $|{\cal I}_n(\ell)|$ denotes the
Koba-Nielsen factor arising from the plane waves of the external vertices whose explicit form is
not relevant for the purpose of this paper but can be looked at in
\refs{\oneloopI,\oneloopII,\oneloopIII}. In addition, $\ell^m = \oint_A dz \Pi^m(z)$ is the
{\it loop momentum} of the chiral splitting formalism \refs{\verlindes,\DHokerPDL,\xerox}
and $A$ denotes the A-cycle of the genus-one
surface under consideration.

\paragraph{Vertex operators} For $n$-point amplitudes involving one first-level massive state
and $(n{-}1)$ massless states, we place the massive leg $\u1$ in the unintegrated massive vertex
and use \BCpaper
\eqn\unint{
V_{\underline 1}= [\l^\a[\p\t^\b B^1_{\a\b}]_0]_0 + [\l^\a[\Pi^m H^m_{1\,\a}]_0]_0
+ 2\ap [\l^\a[d_\b C_1^\b{}_\a]_0]_0
+ \ap[\l^\a[N^{mn} F^1_{\a mn}]_0]_0\,,
}
where the superfields $B_{\a\b}$, $H^m_\a$, $C^\a{}_\b$, $F_{\a mn}$ and $G^{mn} =
-{1\over144}\big[(D\g^m H^n)+(D\g^n H^m)\big]$ describe the open-string massive
supermultiplet at the first mass level. This is composed of the symmetric traceless $g_{mn}$ and the totally
antisymmetric
$b_{mnp}$ bosonic fields and a fermionic field $\psi^m_\a$ comprising
$128+128$ degrees of freedom. They are subject to the transversality constraint
$k^m g_{mn} = k^m b_{mnp} = k_m\psi^m_\a = 0$ \BCpaper.
For the remaining massless integrated vertices $U_i(z_i)$ we use \psf,
\eqn\integratedU{
U_i(z_i) = [\p\t^\a  A^i_\alpha]_0 + [\Pi^m A^i_m]_0
+ 2\ap [d_\a  W_i^\a]_0 + \ap[ N^{mn} F^i_{mn}]_0\,,
}
where $A_\a$, $A^m$, $W^\a$ and $F^{mn}$ are the super-Yang-Mills
superfields of \wittentwistor\ (for a review, see \treereview).
The normal-ordering bracket $[ \ldots]_0$ is reviewed in \UV.

\paragraph{Zero-mode integrations} We refer the reader to the discussion in section 2.1 of \oneloopI\
for more details; the summary is that the zero-mode saturation of the pure
spinor variables imply two different contributions from the external vertices: terms proportional to
$d_\a d_\b N^{mn}$ and terms proportional to $d_\a d_\b d_\g d_\d$. The resulting contribution of
the zero-mode integration can be determined by a group-theory analysis \oneloopI,
\eqnn\efrule
\eqnn\efruletwo
$$\eqalignno{
\int
d_\a d_\b N^{mn}&\rightarrow (\l\g^{[m})_\a(\l\g^{n]})_\b\,, &\efrule\cr
\int
d_\a d_\b d_\g d_\d &\rightarrow
\ell_m (\l\g^{a})_{[\a}(\l\g^{b})_\b(\g^{abm})_{\g\d]}\,, &\efruletwo
}$$
where $\ell_m$ represents the loop momentum
and
the integral sign represents the integration over
the different contributions from the $b$ ghost and
picture-changing operators using
the zero-mode measures of \MPS.

\paragraph{Worldsheet functions}
Two families of worldsheet functions $f^{(n)}(z,\tau)$ and $g^{(n)}(z,\tau)$
indexed by the integer $n$ appearing in one-loop amplitudes are discussed at length in \oneloopII.
In this paper, however, only two such functions make an appearance: the meromorphic
$g^{(1)}(z,\tau)=\p\log\theta_1(z,\tau)$ which captures the simple-pole OPE singularity of a $bc$
system of conformal weights $(1,0)$, where $\theta_1(z,\tau)$ is the odd Jacobi theta function and
the doubly periodic but non-holomorphic $f^{(1)}(z,\tau)=g^{(1)}(z,\tau)+2\pi i (\Im z/\Im \tau)$,
with modular weight one.

\paragraph{Equations of motion and BRST charge} The (linearized) massless $[A_\a,A^m,W^\a,F^{mn}]$ and massive
$[G^{mn},B_{\a\b}, H^m_\a, C^\a{}_\b, F_{\a mn}]$ superfields satisfy the following equations of motion
under the supersymmetric derivative $D_\a = {\p\over \p\t^\a}+\half(\g^m\t)_\a \p_m$,
\eqn\SYMBRST{
\eqalign{
D_\a A_\beta +  D_\beta A_\a &= (\g^m)_{\a\beta}A_m\,,\cr
D_\a A_m &= (\g^m W)_\a + \p_m A_\a\,,\cr
}\qquad
\eqalign{
D_\a W^\b &= {1\over4}(\g^{mn})_{\a}{}^\b F_{mn},\cr
D_\a F_{mn} &=\p_m(\g_n W)_\a- \p_n(\g_m W)_a\,,\cr
}}
and
\eqnn\eomtheta
$$\eqalignno{
D_\a G^{mn} &= -{1\over18}\p_p(\g^{pm}H^n)_\a - {1\over18}\p_p(\g^{pn}H^m)_\a\,, &\eomtheta\cr
D_\a B_{mnp} &= -{1\over18}(\g^{mn}H^p)_\a + {\ap\over18}\p_a\p_m\Big((\g^{an}H^p)_\a
-(\g^{ap}H^n)_\a\Big) + {\rm cyc}(mnp)\,,\cr
D_\a H^m_\b &=-{9\over2}G_{mn}\g^n_{\a\b}-{3\over2}\p_a B_{bcm}\g^{abc}_{\a\b}+{1\over4}\p_a
B_{bcd}\g^{mabcd}_{\a\b}\,,\cr
D_\a C^\g{}_\b &=-{1\over24}(\g^{mnpq})^\g{}_\b\p_m(\g^{np}H^q)_\a\,,\cr
D_\a F_{\b mn}&={3\over 8\ap}B_{amn}\g^a_{\a\b}
+ {1\over32\ap}B_{abc}\g^{mnabc}_{\a\b}\cr
&+{3\over64}\Big[
15B_{amb}\p_n\p_c \g^{cba}_{\a\b}
-{1\over\ap}B_{amb}\g^{nba}_{\a\b}
-3\p_m\p_d B_{abc}\g^{dnbca}_{\a\b}\cr
&\qquad{}-6\p_b G_{am}\g^{bna}_{\a\b}
+ 42\p_n G_{am}\g^a_{\a\b} - (m\leftrightarrow n)
\Big]\cr
}$$
where $B_{mnp}=\g_{mnp}^{\a\b}B_{\a\b}$.
If we define
\eqn\lfields{
\l^\a B_{\a\b}=(\l B)_\b,\quad
\l^\a H^m_\a = (\l H^m)\,,\quad
C^\b{}_\a \l^\a = (C\l)^\b\,,\quad
\l^\a F_{\a mn} = (\l F)_{mn}\,,
}
and use the BRST charge $Q$
\eqn\BRSTcharge{
Q=\l^\a D_\a
}
the equations of motion \eomtheta\ simplify drastically.
In this case we get
\eqn\massQ{
\eqalign{
Q(\l B)_\a &= (\l\g^m)_\a (\l H)_m\,,\cr
Q(\l H^m) &= (\l\g^m C\l)\,,\cr
}\qquad
\eqalign{
Q(C\l)^\a &= {1\over4}(\l\g^{mn})^\a (\l F)_{mn} = {1\over4}(\l\g^{mn})^\a \p_m(\l H_n)\,,\cr
Q(\l F)_{mn} &= {1\over2}\p_{[m}(\l\g_{n]} C\l)
-{1\over16}\p^p(\l\g_{[m} C\g_{n]p}\l)\,,
}
}
Alternatively, the equation of motion
of $(\l F)_{mn}$ can also be written as
\eqn\altQF{
Q(\l F)_{mn} = -{1\over32\ap}(\l\g^{mn})^\b (\l B)_\b
+ {9\over64}\big[\p_n\p_p(\l\g^{mp})^\b(\l B)_\b - (m\leftrightarrow n)\big]\,.
}
One can also show that $\p_m \g^m_{\a\b}C^\b{}_\g = {1\over4\ap}B_{\a\g}$.

\newsubsec\treesec Tree-level amplitudes with one massive state

The open string $n$-point amplitudes at tree level with one first-level massive state and $n{-}1$
massless states were obtained in \massSweden\ in terms of Berends-Giele component currents. Their
pure spinor superspace expressions were subsequently found in \towmass. We briefly review the
pure spinor results below.

\newsubsubsec\tpttreesec Three points

At tree-level, the three-point open-string amplitude of one first-level massive state labelled by
$\u1$ and two massless states labelled $2,3$ is given in pure spinor superspace
by \refs{\UV,\towmass,\PSthreemass}
\eqn\treetptcorr{
\cK_{\u1|2,3}^{\rm tree}=(\l H_1^m)V_2(\l\g^m W_3)\,,
}
where $V_2 = \l^\a A_\a^2$.
Using the equations of motion \SYMBRST\ and \eomtheta\ one can show
that \treetptcorr\ is BRST closed. Computing its component expansion also shows that it is not BRST
exact\foot{It would be BRST exact if the momentum phase space is such that $(k_1+k_2)^2\neq0$ \wip.}
and therefore it is in the cohomology of the BRST charge.

\newsubsubsec\fpttreesec Four points

At tree-level, the four-point open-string amplitude
of one first-level massive state labeled by $\u1$ and three massless states labeled by $2,3,4$
can be written in terms of its kinematic factor in pure spinor superspace \towmass\foot{See \massSweden\
for the full string amplitude, including the Beta function.},
\eqn\treefpt{
\cK_{\u1|2,3,4}^{\rm tree}= (\l H_1^m)C^m_{2|34}\,.
}
In the above, $C^m_{2|34}$ denotes the BRST-closed combination \towmass
\eqn\Cmtree{
C_{2|34}^m = M_{23}(\l\g^m \cW_{4})
+ M_{2}(\l\g^m \cW_{34})
- M_{24}(\l\g^m \cW_{3})\,,
}
where $M_P$ and $\cW^\a_P$ are the multiparticle Berends-Giele currents of the multiparticle
superfields $V_P$ and $W^\a_P$, see \treereview\ for a review.

\newnewsec\massBB Massive multiparticle superfields

It is clear that the massless unintegrated vertex given by
$V_1 = \l^\a A^1_\a(x,\t)$ does not contribute any $d_\a$ or $N^{mn}$ zero modes simply because
these variables are absent in the vertex. However, this
is no longer the case when the unintegrated vertex represents a {\it massive} string state. The
consequence is that all vertices in $V_\u1 U_2 \ldots U_n$ can possibly contribute those zero modes.
Comparing the unintegrated massive vertex \unint\ with the massless integrated vertex \integratedU,
it is easy to see that $(\l H^m_1)$, $(C_1\l)^\a$ and $(\l F_1)_{mn}$
play an analogous role as
the super-Yang-Mills superfields $A^m_i$, $W_i^\a$ and $F_i^{mn}$,
\eqn\analogous{
A^m \leftrightarrow (\l H^m),\quad
W^\a \leftrightarrow (C\l)^\a,\quad
F_{mn} \leftrightarrow (\l F)_{mn}\,.
}
We will use
this observation in the construction of BRST-covariant objects below.

When the saturation of zero modes admits a prior OPE contraction among the vertices, their
contribution is summarized by {\it multiparticle} superfields. They are defined as the coefficients of the remaining
conformal weight-one
variables in a suitable pole of the OPE. When the vertices involved in the OPE are massless, this
is encoded in the
massless multiparticle superfields of \EOMbbs\ (for a review see \treereview). In the present
case, the OPE may involve the massive vertex $V_\u1$.
For example, a multiparticle superfield $(C_{12}\l)^\a$ can be
read off from the coefficient of
$d_\a$ in the OPE bracket $[V_\u1 U_2]_1$,
\eqnn\Cuddef
$$\eqalignno{
(C_{12}\l)^\a &=
- (\l H_1^{m})W_2^\a ik_2^m
- (C_1\l)^\a( ik_1\cdot A_2)
+ \big(D_\b(C_1\l)^\a\big) W_2^\b\cr
&-{1\over4}\big[(\g^{mn}C_1\l)^\a F_2^{mn}- (C_1\g^{mn}\l)^\a F_2^{mn}\big]\,. &\Cuddef\cr
}$$
After contraction with $(\l\g^m)_\a$, its BRST variation can be shown to be
\eqnn\QCdef
$$\eqalignno{
Q(\l \g^m C_{12}\l) &= s_{12}V_2(\l\g^m C_1)\,, &\QCdef\cr
}$$
which has the desired properties for our purposes, see below.
Unfortunately, the definition of the multiparticle superfield $(\l F_{12})_{mn}$ is not so
straightforward as reading off the coefficient of $N^{mn}$ in the OPE $[V_\u1 U_2]_1$.
The constraint identity $[N^{mn}\l^\b]_0\g^m_{\b\g}  =
\half[J\l^\b]_0\g^n_{\b\g} +
2(\g^n \p\l)_\g$ leads to a non-unique definition of
the coefficient of $N^{nm}$. Luckily, BRST covariance can be used to the rescue and we
define $(\l F_{12})_{mn}$ such that
\eqn\QFdef{
Q\big[(\l F_{12})_{mn}(\l\g^m W_3)(\l\g^n W_4)\big] =
s_{12}V_2(\l F_{1})_{mn} (\l\g^m W_3)(\l\g^n W_4)\,.
}
A solution to \QFdef\ is given by
\eqn\Fuddef{
(\l F_{12})_{mn} = -\big(D_\a (\l F_1)_{mn}\big)W_2^\a - (\l F_1)_{mn}(ik_1\cdot A_2)
-(\l\g^{pq}F_1)_{mn}F^2_{pq}\,.
}
In doing the above calculations, it is necessary  to use
that there is no $4$-form irrep in the decomposition of $\l^3 W$ \LiEprogram\ to conclude
$(\l\g^{[m}W) (\l\g^{npqrs]}\l)=0$. In particular
\eqnn\Canonid
$$\eqalignno{
(\l \g^{q}  W^4) B^{1}_{mnp}A^{2}_{q} (\l \g^{mnpk^{1}k^{2}} \l)   &=
-  (\l \g^{k^{1}}  W^4) B^{1}_{mnp}A^{2}_{q} (\l \g^{qmnpk^{2}} \l) &\Canonid\cr&
 +  (\l \g^{k^{2}}  W^4) B^{1}_{mnp}A^{2}_{q} (\l \g^{qmnpk^{1}} \l) \cr&
 +  3 (\l \g^{m}  W^4) B^{1}_{mnp}A^{2}_{q} (\l \g^{qnpk^{1}k^{2}} \l)
}$$
\paragraph{Scalar BRST blocks} Since the massless
$W^\a$ and $F^{mn}$
appear in schematic form as $(\l\g^m W)(\l\g^n W)F_{mn}$ as the result of the zero-mode
integration \efrule, we can use the observation \analogous\ to
define massive superfield building blocks
that
are analogous to the massless $T_{A,B,C}$ in \oneloopI
\eqnn\Tdefs
$$\eqalignno{
T_{\underline 1,2,3}&=
- (\l\g^m C_1\l)(\l\g^n W_2)F_3^{mn}
- (\l\g^m C_1\l)F_2^{mn}(\l\g^n W_3) &\Tdefs\cr&
- (\l F_1)_{mn}(\l\g^m W_2)(\l\g^n W_3)\cr
T_{\underline 1 2,3,4} &= -(\l\g^m C_{12}\l)(\l\g^n W_3)F_4^{mn}
- (\l\g^m C_{12}\l)F_3^{mn}(\l\g^n W_4) \cr
&- (\l F_{12})_{mn}(\l\g^m W_3)(\l\g^n W_4)\cr
T_{\underline 1 ,23,4} &=
- (\l\g^m C_{1}\l)(\l\g^n W_{23})F_4^{mn}
- (\l\g^m C_{1}\l)F_{23}^{mn}(\l\g^n W_{4})\cr
&- (\l F_{1})_{mn}(\l\g^m W_{23})(\l\g^n W_4)\,.
}$$
Their BRST variations can be readily computed to be
\eqnn\QTdefs
$$\eqalignno{
QT_{\underline 1,2,3} &=0 &\QTdefs\cr
Q T_{\underline 12,3,4} &= s_{12}V_2T_{\underline1,3,4}\cr
Q T_{\underline 1,23,4} &= s_{23}\big(V_3T_{\underline1,2,4} - V_2T_{\underline1,3,4}\big)\,.\cr
}$$
The scalar BRST blocks $T_{\u1 i,j,k}$ and $T_{\u1,ij,k}$ for different labels $i,j,k$ are
obtained from relabeling the expressions above.
\paragraph{Vectorial BRST block}
Similarly,
the zero-mode integration of four $d_\a$ given in \efruletwo\ results
in a vector constructed of
three massless $W^\a$ and one massive $(C\l)^\a$. Inspired by the massless expression of
$W^m_{A,B,C,D}$ given in
\oneloopI\ and the correspondence \analogous, we propose
\eqnn\Wmdef
$$\eqalignno{
W^m_{\underline 1,2,3,4} & =
 -  {5 \over 12}\, (\l \g^{a}  C_1\l) (\l \g^{b}  W^2) (W^{3} \g^{mab}  W^4) &\Wmdef\cr&
 -  {5 \over 12}\, (\l \g^{a}  C_1\l) (\l \g^{b}  W^4) (W^{2} \g^{mab}  W^3)\cr&
 -  {5 \over 12}\, (\l \g^{a}  C_1\l) (\l \g^{b}  W^3) (W^{4} \g^{mab}  W^2)\cr&
 -  {1 \over 4}\, (\l \g^{a}  W^2) (\l \g^{b}  W^3) (W^{4} \g^{mab}  C_1\l)\cr&
 -  {1 \over 4}\, (\l \g^{a}  W^4) (\l \g^{b}  W^2) (W^{3} \g^{mab}  C_1\l)\cr&
 -  {1 \over 4}\, (\l \g^{a}  W^3) (\l \g^{b}  W^4) (W^{2} \g^{mab}  C_1\l)\,.
}$$
The
relative coefficients in \Wmdef\ are determined from the requirement of BRST covariance; it follows
from the massive and massless superfield equations of motion that
\eqn\QWndef{
QW^m_{\underline 1,2,3,4} = - (\l\g^m C_1\l)T_{2,3,4}-\big[(\l\g^m W_2)T_{\underline 1,3,4}
+ (2\leftrightarrow3,4)\big]\,.
}
As discussed in \oneloopI, considerations of BRST covariance involving a mixture of left- and
right-movers require a vectorial object whose
BRST variation carries the vector index exclusively in momenta. Inspired by the
massless case of \oneloopI, one is led to consider the following vectorial BRST block
\eqn\Tmdef{
T^m_{\underline 1,2,3,4}=-(\l H_1^m)T_{2,3,4}-\big[A_2^m T_{\underline 1,3,4} +
2\leftrightarrow3,4\big]- W^m_{\underline 1,2,3,4}
}
with BRST variation
\eqn\QTmdef{
QT^m_{\underline 1,2,3,4} =  -ik_2^m V_2 T_{\underline 1,3,4} + (2\leftrightarrow3,4)\,.
}
\paragraph{Berends-Giele}
For convenience, we introduce the Berends-Giele non-local counterparts of the above BRST blocks as
\eqnn\BGcurr
$$\displaylines{
M_{\u12,3,4} = {1\over s_{12}}T_{\u12,3,4},\quad
M_{\u1,23,4} = {1\over s_{23}}T_{\u1,23,4},\quad
M_{\u13,2,4} = {1\over s_{13}}T_{\u13,2,4},\hfil\BGcurr\hfilneg\cr
M^m_{\u1,2,3,4} = T^m_{\u1,2,3,4}\,,\quad M_{\u1,2,3}=T_{\u1,2,3}
}$$
which satisfy
\eqnn\QBGcurr
$$\displaylines{
QM_{\u12,3,4} = M_2M_{\u1,3,4},\;
QM_{\u13,2,4} = M_3M_{\u1,2,4},\;
QM_{\u1,23,4} = M_3M_{\u1,2,4}-M_2M_{\u1,3,4},\hfil\QBGcurr\hfilneg\cr
QM^m_{\u1,2,3,4} = - ik_2^m M_2 M_{\u1,3,4}+(2\leftrightarrow3,4),\quad QM_{\u1,2,3} = 0\,.
}$$

\newsec{The massive 1-loop amplitude}

In this section we will use the multiparticle massive superfields defined in section~\massBB\ to
construct a single-valued and BRST-invariant expression for the four-point open string amplitude at one loop
involving one first-level massive state. As a warm-up we will derive, for the first time, the
three-point massive amplitude at one loop.

\newsubsec\tptampsec Three points

Using the pure spinor prescription \onepresc\ with a massive $V_\u1$ yields a unique saturation of
the zero modes $d_\a d_\b N^{mn}$ as there is only three external vertices. More precisely,
\eqn\tolp{
\Big[V_\u1 U_2 U_3\Big]_{ddN} = d_\a d_\b N^{mn}\Big((C_1\l)^\a W_2^\b F_3^{mn}
+ (C_1\l)^\a F_2^{mn} W_3^\b
+ (\l F_1)_{mn}W_2^\a W_3^\b\Big).
}
In particular, there are not enough external vertices to produce a contribution with the loop
momentum, so the three-point correlator defined in \theamp\ does not depend on $\ell^m$.
Integrating out $d_\a d_\b N^{mn}$ using \efrule\ yields
\eqn\tptcorrtmp{
\cK_3 =
- (\l\g^m C_1\l)(\l\g^n W_2)F_3^{mn}
- (\l\g^m C_1\l)F_2^{mn}(\l\g^n W_3)
- (\l F_1)_{mn}(\l\g^m W_2)(\l\g^n W_3).
}
As will be demonstrated in \Tuid, this becomes
\eqn\tptcorr{
\cK_3 = (\l F_1)_{mn}(\l\g^m W_2)(\l\g^n W_3)\,.
}
The
component
expansion of \tptcorr\ is straightforward to calculate \refs{\PSS,\FORM} using the pure spinor bracket $\langle
(\l\g^m\t)(\l\g^n\t)(\l\g^p\t)(\t\g_{mnp}\t)\rangle=1$ \psf\ and the theta expansions of the various
superfields \refs{\HarnadS,\treereview,\masstheta,\towmass}. It yields
\eqn\threecomp{
\langle \cK_3 \rangle = {1 \over 640 \ap}\Big(
g_{1\,mn} f_2^{ma}f_3^{na}
-{1\over2\ap}b_{1\,mnp}f_2^{mn}e_3^p
\Big)+\hbox{ fermions}
}
where $g_{1\,mn}$ and $b_{1\,mnp}$ are the bosonic massive polarizations \BCpaper,
and $f_j^{mn}=ik_j^m e_j^n - ik_j^n e_j^m$ are the linearized field strengths and $e_i^m$ are the gluons.

In the appendix~\proofap, we will use BRST cohomology manipulations similar to the ones used in \mafraids\ 
to show the relation between the one-loop correlator \tptcorr\ and tree-level amplitude \treetptcorr:
\eqn\lemma{
\langle (\l F_1)_{mn}(\l\g^m W_2)(\l\g^n W_3)\rangle =
{1\over 2\ap}\langle (\l H_1^m)V_2(\l\g^m W_3)\rangle={1\over2\ap}A(\u1,2,3)\,,
}
where $1/2\ap = - (k_2\cdot k_3)$.
For convenience, we define the BRST invariant
\eqn\Ctdef{
C_{\u1|2,3} = T_{\u1,2,3}\,,
}
and write the one-loop three-point correlator as $\cK_3 = C_{\u1|2,3}$.

\newsubsec\fptampsec Four points

The four-point amplitude with one massive state admits two types of zero-mode saturation:
contributions with
$d_\a d_\b N^{mn}$ or $d_\a d_\b d_\g d_\d$. The first kind leads to OPE contractions among
the vertices leading to the massive multiparticle superfields discussed in section~\massBB\ as
well as $\ell_m (\l H^m_1)T_{2,3,4}$ or $\ell_m A^m_2 T_{\u1,3,4} + (2\leftrightarrow3,4)$.
The zero-mode integration of the second kind of contribution leads to $\ell_m W^m_{\u1,2,3,4}$.
These various contributions are organized according to a desired property of BRST covariance of
the individual blocks of superfields resulting in an overall BRST-invariant correlator. As the
massive four-point amplitude is analogous to the massless five-point amplitude, the five-point
correlator of \oneloopIII\ leads to the following proposal,
\eqnn\fpto
$$\eqalignno{
\cK_4(\ell) &=
\ell_m T^m_{\underline 1,2,3,4}
+ g^{(1)}_{12}T_{\underline 12,3,4}
+ g^{(1)}_{13}T_{\underline 13,2,4}
+ g^{(1)}_{14}T_{\underline 14,2,3}&\fpto\cr
&+ g^{(1)}_{23}T_{\underline 1,23,4}
+ g^{(1)}_{24}T_{\underline 1,24,3}
+ g^{(1)}_{34}T_{\underline 1,34,2}\,.
}$$
Its BRST variation is a total worldsheet derivative,
\eqn\totder{
Q\cK_4(\ell) = V_2T_{\u1,3,4}\Big(-(ik_2\cdot\ell) - s_{21}g_{21}^{(1)} - s_{23}g_{23}^{(1)} -
s_{24}g_{24}^{(1)}\Big) + (2\leftrightarrow3,4)\,.
}
To see this, one uses the worldsheet derivative of the Koba-Nielsen factor \oneloopII,
\eqn\zderiv{
{\p\over\p z_i}\cI_n(\ell) =\big(
\ell \cdot ik_i + \sum_{j\neq i}^n s_{ij} g^{(1)}_{ij}\big)\cI_n(\ell)
}
is proportional to the right-hand side of \totder.
Therefore, once the integration over the worldsheet insertion
points is carried out, the amplitude \theamp\ is BRST invariant as the boundary terms vanish
using the canceled propagator argument.

\newsubsubsec\brstsec Manifesting BRST invariance

Similarly to the discussion of BRST invariance in the massless one-loop amplitudes
of \oneloopIII, one can use the integration by parts identities
\eqnn\ibps
$$\eqalignno{
s_{12}g_{12}^{(1)}&\sim (\ell.ik_2) + s_{23}g_{23}^{(1)}+s_{24}g_{24}^{(1)}, &\ibps\cr
s_{13}g_{13}^{(1)}&\sim (\ell.ik_3) - s_{23}g_{23}^{(1)}+s_{34}g_{34}^{(1)},\cr
s_{14}g_{14}^{(1)}&\sim (\ell.ik_4) - s_{24}g_{24}^{(1)}-s_{34}g_{34}^{(1)},
}$$
to obtain the BRST-invariant correlator
\eqn\Bcorr{
\cK_4(\ell) \sim \ell_m C^m_{\u1|2,3,4}
+ s_{23}g_{23}^{(1)}C_{\u1|23,4}
+ s_{24}g_{24}^{(1)}C_{\u1|24,3}
+ s_{34}g_{34}^{(1)}C_{\u1|34,2}\,.
}
In the above we defined the scalar BRST invariants
\eqnn\Cmudtqdef
\eqnn\Cudtqdef
\eqnn\Cudqtdef
\eqnn\Cutqddef
$$\eqalignno{
C^m_{\u1|2,3,4} &= M^m_{\u1,2,3,4}
+ ik_2^m M_{\u12,3,4}+ik_3^m M_{\u13,2,4}+ik_4^m M_{\u14,2,3}
&\Cmudtqdef\cr
C_{\u1|23,4} &= M_{\u12,3,4}+M_{\u1,23,4}-M_{\u13,2,4} &\Cudtqdef\cr
C_{\u1|24,3} &= M_{\u12,3,4}+M_{\u1,24,3}-M_{\u14,2,3} &\Cudqtdef\cr
C_{\u1|34,2} &= M_{\u13,2,4}+M_{\u1,34,2}-M_{\u14,2,3}, &\Cutqddef
}$$
in terms of the Berends-Giele currents defined in \BGcurr. Using the
BRST variations \QBGcurr, one can show that the above combinations are BRST closed,
\eqn\Cclosed{
QC^m_{\u1|2,3,4} = QC_{\u1|23,4} = QC_{\u1|24,3} = QC_{\u1|34,2} = 0.
}
The above four-point BRST invariants with one massive state have the same
structure as the five-point BRST invariants in the massless
one-loop amplitudes of \refs{\oneloopI,\oneloopIII}.

\newsubsubsec\homsec Manifesting single valuedness

It turns out that
the massive BRST invariants \Cmudtqdef-\Cutqddef\ obey the analog momentum-contraction identities as their massless
counterparts,
i.e.,
\eqnn\Cmid
$$\eqalignno{
ik^m_2 \langle C^m_{\u1|2,3,4}\rangle
- s_{23}\langle C_{\u1|23,4}\rangle
- s_{24}\langle C_{\u1|24,3}\rangle &= 0,&\Cmid\cr
ik_1^m \langle C^m_{\u1|2,3,4}\rangle &= 0
}$$
As discussed in \oneloopIII, these identities are sufficient to show that
the correlator \Bcorr\ is single-valued and can be rewritten in a manifestly single-valued
manner as
\eqn\BcorrSV{
\cK_4 =
s_{23}f^{(1)}_{23}C_{\u1|23,4}
+ s_{24}f^{(1)}_{24}C_{\u1|24,3}
+ s_{34}f^{(1)}_{34}C_{\u1|34,2}\,,
}
after integrating out the loop momentum.

\newsubsubsec\comprel Relation to the four-point tree amplitude

We know that the pure spinor superspace expressions of one-loop and tree-level correlators are
proportional in the massive three-point \lemma\ and massless four-point \mafraids\ cases.
It is natural to ask what happens in the case of massive four-point correlators. Indeed,
one can show directly via their component expansions that
(note $s_{ij}=(k_i\cdot k_j)$)
\eqnn\CAtreeid
$$\eqalignno{
\langle C_{\u1|23,4}\rangle &=
s_{24}s_{34}\Big(
{1\over s_{12}}
+{1\over s_{13}}
\Big)
\cK^{\rm tree}_{\u1|2,3,4} &\CAtreeid\cr
\langle C_{\u1|24,3}\rangle &=
-s_{23}s_{34}\Big(
{1\over s_{12}}
+{1\over s_{14}}
\Big)
\cK^{\rm tree}_{\u1|2,3,4}\cr
\langle C_{\u1|34,2}\rangle &=
s_{23}s_{24}\Big(
{1\over s_{13}}
+{1\over s_{14}}
\Big)
\cK^{\rm tree}_{\u1|2,3,4}\cr
}
$$
where the tree-level kinematic factor is given by \treefpt\ and the one-loop BRST invariant is given by
\Cudtqdef.
Note that \CAtreeid\ has been checked to hold for both  bosonic massive states in the first mass
level ($g_{mn}$ and $b_{mnp}$). By supersymmetry, also
the fermionic components will match on both sides.
Similar identities were found with the RNS formalism in \wildenauer, but the
analysis was restricted to the bosonic state $g_{mn}$ (where they agree).

The above calculations use the convention
$s_{ij}=k_i\cdot k_j$, and therefore
\eqn\momps{\eqalign{
s_{13} &= {1\over2\ap}-s_{12}-s_{23},\cr
s_{14} &= {1\over2\ap}+s_{23}
}\qquad
\eqalign{
s_{24}&=-{1\over2\ap}+s_{13}
=-s_{12}-s_{23},\cr
s_{34}&=-{1\over2\ap}+s_{12}
}}
where $k_1^2 = -1/\ap$ and $k_2^2= k_3^2= k_4^2=0$.
In addition, the relations
\eqn\ugly{\eqalign{
{s_{12}\over s_{13}}&={1\over2\ap s_{13}}-{s_{23}\over s_{13}} -1\cr
{s_{23}\over s_{14}}&=1-{1\over2\ap s_{14}}\cr
}\qquad\quad\eqalign{
{s_{12}\over s_{24}}&=-1-{s_{23}\over s_{24}}\cr
{s_{12}\over s_{34}}&={1\over2\ap s_{34}}+1\cr
}\qquad\quad\eqalign{
{s_{23}\over s_{13}s_{34}}&=-{1\over s_{23}}-{1\over s_{34}}\cr
{s_{23}\over s_{12}s_{24}}&=-{1\over s_{12}}-{1\over s_{24}}\cr
}
}
follow from \momps\ and are useful in checking \CAtreeid.

\newnewsec\closedsec Closed strings

We have seen in \tptcorr\ and \BcorrSV\ that the {\it open-string} one-loop three- and four-point
correlators with one massive state
can be written in terms of tree-level kinematic factors.
This section is motivated by the desire to know to which extent the {\it closed-string}
four-point\foot{From \tptcorr\ it immediately follows
that the holomorphic square of the open-string three-point correlator can be written
in terms of tree-level amplitudes.} correlator admits a rewriting in terms of tree-level
kinematic data, if at all or for which combination of massive and massless closed-string states.

\newsubsec\fptcorr The four-point correlator

According to the chiral splitting formalism \refs{\verlindes,\DHokerPDL,\xerox}, the correlator $M_4$
of the closed string amplitude
is obtained after integrating the loop momentum in the holomorphic square
of the open-string correlator \Bcorr. Using the techniques described in \oneloopIII\ this leads to
the four-point closed-string correlator $M_4$
\eqn\closedell{
M_4 = |s_{23}f_{23}^{(1)}C_{\u1|23,4} + (2,3|2,3,4)|^2 + {\pi\over
\Im\tau}C^m_{\u1,2,3,4}\tilde C^m_{\u1,2,3,4}\,.
}
The four-point closed string amplitude at genus-one is then written as
\eqn\fourclosed{
{\cal M}_4 = \int_{\cal F}d^2\tau\int d^2z_2 d^2z_3 d^2z_4 \hat{\cal I}_4 M_4
}
where ${\cal F}$ is the fundamental domain of the genus-one moduli space and $\hat{\cal I}_4$ is
the genus-one Koba-Nielsen factor \oneloopIII\


\newsubsec\casop One-loop versus tree-level

The form of the correlator \closedell\ featuring two distinct contributions, one with a
holomorphic square and another with a vectorial contraction between
the left- and right-movers, is reminiscent of the one-loop correlator for five {\it massless}
closed-string states considered in \oneloopMichael. In that case, the analogous vectorial
contraction (see their equation (3.35)) was shown to be rewritten in terms of the massless
five-point tree-level amplitudes for type IIB states. Moreover, that analysis had
an immediate impact on the S-duality symmetry of type IIB strings.
While the massive amplitudes considered in this paper have no direct relation to the S-duality
property of massless amplitudes,
it is still interesting to check whether the massive correlator \closedell\ at one-loop can be written in terms
of its massive counterpart \treefpt\ at tree level\foot{I thank Oliver
Schlotterer for raising this question.}.
Surprisingly, we will show below that, for certain combinations of massive and massless states,
this is indeed possible.

\paragraph{First-level massive closed-string spectrum} The closed-string states in the first mass
level of bosonic amplitudes can be characterized by the dimension of their $SO(9)$ irrep:
\eqn\cstates{
\bf1,\bf36,\bf44,\bf84,\bf126,\bf231,\bf450,\bf495,\bf594,
\bf910,\bf924,\bf1980,\bf2457,\bf2772.
}
An explicit component expansion using the closed-string state
decompositions listed in the Appendix~\decapp, reveals that the left-right contraction
term in \closedell\ can be written in terms of the tree-level amplitude \treefpt,
\eqn\Sdualmass{
s_{12}s_{13}s_{14}C^m_{\u1|2,3,4}\tilde C^m_{\u1|2,3,4} = 4s^2_{23}s_{24}^2 s_{34}^2
\cK^{\rm tree}_{\u1|2,3,4}\tilde\cK^{\rm tree}_{\u1|2,3,4}\,,
}
for the following combination of external states in type IIB
\eqnn\listst
$$\eqalignno{
g h^3, g\phi h^2:&\qquad g\in\{\bf1,\bf44,\bf84,
\bf450, \bf495, \bf1980, \bf2457\} &\listst\cr
g h^2 \phi:&\qquad g\in\{\bf126,\bf231\}\cr
g \phi^3:&\qquad g\in\{\bf231\}\cr
g b\phi^2:&\qquad g\in\{\bf2772\}\cr
}$$
where the massive state is represented by $g$, the graviton by $h$, the NS-NS $2$-form by $b$ and the
dilaton by $\phi$. Similarly, equation \Sdualmass\ is satisfied in type IIA for\foot{A natural
question to ask, in view of the results of \oneloopMichael, is whether different states satisfy
a modified version of \Sdualmass\ where the relative coefficient $4$ is replaced by a rational
number. However, we have not found such a case.}
\eqnn\liststA
$$\eqalignno{
g h^3, gh\phi^2:&\qquad g\in\{\bf84,\bf450,\bf2457\} &\liststA\cr
g h^2\phi :&\qquad g\in\{\bf126,\bf231\}\cr
g\phi^3 :&\qquad g\in\{\bf231\}\cr
}$$
Note that the type IIB result \listst\
for the states $\bf44$ and $\bf495$ is only achieved if their decompositions coming from
$g_{mn}\otimes\tilde g_{pq}$ and $b_{mnp}\otimes\tilde b_{qrs}$
are fine-tuned with a relative coefficient. More explicitly,
for $\bf44$,
\eqnn\gtwolc
$$\eqalignno{
g_{mn}\otimes\tilde g_{pq} &= \Pi^{mn}_{am_1}\Pi^{pq}_{am_2}g_{20000}^{m_1m_2},&\gtwolc\cr
b_{mnp}\otimes\tilde b_{qrs} &= {13\over2}\ap\Pi^{mnp}_{abm_1}\Pi^{pqr}_{abm_2}g_{20000}^{m_1m_2}
}$$
while for
$\bf495$ we have,
\eqnn\gztlc
$$\eqalignno{
g_{mn}\otimes\tilde g_{pq} &=
        g_{02000}^{mpqn}+g_{02000}^{npqm}
        + g_{02000}^{mqpn}+g_{02000}^{nqpm},&\gztlc\cr
        b_{mnp}\otimes\tilde b_{qrs} &=-{27\over5}\ap\Pi^{mnp}_{am_1m_2}\Pi^{pqr}_{am_3m_4}g_{02000}^{m_1m_2m_3m_4}
}$$
see the
appendix~\decapp\ for the definitions and the complete list.

It is interesting to note that a linear combination of massive
first-level states
in the scalar representation $\bf1$ was considered in \minahan. However the
scalar $\bf1$ in $\bf44\otimes\bf44$ and the scalar $\bf1$ in $\bf84\otimes\bf84$ individually
satisfy \Sdualmass\ in the type IIB theory when paired with $h^3$ or $\phi^2 h$, so no linear combination of the scalars
is necessary in this case.

\newnewsec\conc Conclusions

In this paper we used arguments based on pure spinor BRST covariance and single-valuedness
to propose expressions for
one-loop correlators of the open string for three and four points when one of the external legs is
a first-level massive state.

We then integrated the loop momentum to obtain the closed-string correlator \closedell\ of the one-loop
four-point massive amplitude. Its similarity with the massless
five-point one-loop amplitude in \oneloopMichael\ led us to consider whether
it can be rewritten in terms of tree-level massive kinematic factors. We found certain
combinations of massive and massless states
in \listst\ for the type IIB and \liststA\ for the type IIA string theory
for which this rewriting can be done.
It is worth noting that the analogous analysis of the closed-string five-point
massless one-loop correlator \oneloopMichael\ had implications for the S-duality of type IIB string
theory. It is unclear at the moment whether the results of section~\fptcorr\ are an indication of
something else than a curiosity.
It will be interesting to see
if similar results hold for higher-point and/or higher-loop amplitudes.
We leave these investigations for the future.

\bigskip
\noindent{\bf Acknowledgements:} I thank Oliver Schlotterer for asking a question which provided
the motivation for this work, for pointing out
the RNS calculations of \wildenauer, and for suggesting the closed-string analysis.
I also thank him for comments on the draft.

\appendix{A}{Proof of tree-level and one-loop relation at three points}
\applab\proofap

\proclaim Lemma. In on-shell pure spinor superspace we have
\eqn\lem{
\langle (\l F_1)_{mn}(\l\g^m W_2)(\l\g^n W_3)\rangle
= {1\over2\ap}\langle(\l H_1^m)V_2(\l \g^m W_3)\rangle\,.
}

\noindent{\it Proof.} To see this, we start from the expression \tptcorr\ and observe that $(\l\g^m C_1\l)=Q(\l H_1^m)$
\towmass. Integrating the BRST charge by parts, using $QF^{mn} = \p^m(\l\g^n W)-\p^n(\l\g^m W)$
and the Dirac equation, leads to
\eqnn\idtmpu
$$\displaylines{
- (\l\g^m C_1\l)(\l\g^n W_2)F_3^{mn}
- (\l\g^m C_1\l)F_2^{mn}(\l\g^n W_3) =\hfil\idtmpu\hfilneg\cr
-\big(\p^n_1(\l H_1^m)-\p^m_1(\l H_1^n)\big) (\l\g^m W_2)(\l\g^n W_3)\,.
}$$
On the other hand, in the Berkovits-Chandia gauge \BCpaper\ we have
\eqn\FBC{
(\l F_1)_{mn}=
{1\over 16}\Big(7\p_m(\l H^n_1)-7\p_n(\l H^m_1) + (\l \g^{k^1 m}H^n_1) - (\l \g^{k^1
n}H^m_1)\Big)\,,
}
and $(\l H^{k^1}_1) = 0$.
After a short calculation using the pure spinor constraint, this implies
\eqn\laFmnId{
2(\l F_1)_{mn}(\l\g^m W_2)(\l\g^n W_3)
= \big(\p^m_1(\l H_1^n)-\p^n_1(\l H_1^m)\big)(\l\g^m W_2)(\l\g^n W_3)
}
Plugging this into \idtmpu\ yields
\eqn\ptid{
- (\l\g^m C_1\l)(\l\g^n W_2)F_3^{mn}
- (\l\g^m C_1\l)F_2^{mn}(\l\g^n W_3) = 2(\l F_1)_{mn}(\l\g^m W_2)(\l\g^n W_3)\,,
}
which means that
\eqn\Tuid{
T_{\u1,2,3} = (\l F_1)_{mn}(\l\g^m W_2)(\l\g^n W_3)\,.
}
Therefore,
\eqnn\firstl
$$\eqalignno{
T_{\u1,2,3} &= (\l F_1)_{mn}(\l\g^m W_2)(\l\g^n W_3)\cr
&= \half\Big(\p^m_1(\l H_1^n)-\p^n_1(\l H_1^m)\Big)(\l\g^m W_2)(\l\g^n W_3)\cr
&= {1\over2}(\l H_1^m)(\l\g^m W_2)\p_2^n(\l\g^n W_3) &\firstl\cr
&- {1\over2}(\l H_1^n)\p_3^m(\l\g^m W_2)(\l\g^n W_3)
}$$
where we used momentum conservation $\p^m_1 = -\p_2^m-\p^m_3$ and the Dirac equation to arrive at
the third equality. Using
$(\l\g^n W_3) = QA^n_3 - \p^n_3 V_3$ in the third line of \firstl\ and
$(\l\g^m W_2) = QA^m_2 - \p^m_2 V_2$ in the fourth yields
\eqnn\Tusec
$$\eqalignno{
T_{\u1,2,3} &= {1\over2}(\p_2\cdot \p_3)(\l H_1^m)\Big(V_2(\l\g^m W_3)+V_3(\l\g^m W_2)\Big)&\Tusec\cr
&= {1\over 2\ap}(\l H_1^m)V_2(\l\g^m W_3)
}$$
where we dropped BRST-exact terms such as $(\l H_1^m)(\l\g^m W_2)\p^n_2QA_3^n$,
used equation (3.9) of \UV\ showing that the two terms inside the parenthesis in
\Tusec\ are equal in the BRST cohomology, and $(\p_2\cdot \p_3) = -
s_{23}={1\over 2\ap}$. Finally, this means
\eqn\lempro{
\langle (\l F_1)_{mn}(\l\g^m W_2)(\l\g^n W_3)\rangle = {1\over 2\ap}\langle(\l H_1^m)V_2(\l\g^m W_3)\rangle
}
or equivalently,
$\langle K^{\rm 1loop}_{100}\rangle = {1\over2\ap}\langle K^{\rm tree}_{100}\rangle$,
finishing the proof. \qed

\appendix{B}{Tensorial description of first-level massive closed-string states}
\applab\decapp

\noindent In this appendix we briefly review the {\it tensorial} description of irreps of
$SO(9)$ and $SO(10)$ using the mapping between Dynkin labels and Young diagrams \fischler\ (see
also \adynkras).
We also review the symmetries of the associated tensors that are not manifestly encoded in
the Young diagrams \refs{\hansenII,\stahngarnir}. Finally, we list the explicit tensorial expressions
of how the closed-string states at the first massive level
are constructed from the holomorphic square of open-string states.

\def\youngdim{0.5}
\paragraph{$SO(9)$ tensor irreps} The irreps of $SO(9)$ are labelled by four Dynkin labels $(a_1 a_2 a_3 a_4)$.
They correspond to a tensor (as opposed to a spinor) when $a_4$ is {\it even}. In this case,
the irrep $(a_1 a_2 a_3 a_4)$ encodes a Young diagram with $a_i$
columns with $i$ boxes for $i=1,2,3$ and $a_4/2$ columns with 4 boxes.
For example $(2200)$ maps to $\yng(4,2)$
because there are two columns with one box and two columns with two boxes.

\paragraph{SO(10) tensor irreps} The irreps of SO(10) labelled by five Dynkin labels $(a_1 a_2
a_3 a_4 a_5)$ correspond to a tensor when $a_4+a_5$ is {\it even}. Their mapping
to Young diagrams works the same way as in $SO(9)$ for the labels $a_1,a_2,a_3$, while the mapping
of the labels
$a_4$ and $a_5$ to columns require a separate analysis depending on two cases:
\smallskip
\item{1.} $a_4\le a_5$: there are $a_4$ columns of $4$ boxes and $(a_5-a_4)/2$ columns of $5$ boxes
\item{2.} $a_4> a_5$: this is the conjugate representation,
there are $a_5$ columns of $4$ boxes and $(a_4-a_5)/2$ columns of $5$ boxes
\smallskip
\noindent For example, $(00011)$ is a tensor that maps to the $4$-form $\yng(1,1,1,1)$.
\def\youngdim{0.3}
Similarly,
both $(00002)=\yng(1,1,1,1,1)$ and $(00020)=\yng(1,1,1,1,1)$ constitute $5$-forms
(self-dual and anti-self-dual).

\paragraph{Young tableaux and tensors} We associate traceless tensors to Young tableaux of $SO(n)$
by populating the columns with the tensor indices according to the antisymmetric basis scheme
\hansenII. For example,
\def\youngdim{1.4}
\eqn\tabtotensor{
\young(1578,26,3,4) \raise12pt\hbox{$\qquad\leftrightarrow\qquad T_{12345678} = T_{[1234][56](78)}$}
}
where the labels $j$ in the tensor are a shorthand for vector indices $m_j$.
In \tabtotensor\ we also display the explicit (anti)symmetries of the associated tensor.
Being traceless irreps of $SO(n)$ means that the trace
with respect to any two indices vanish
\eqn\tracelessdef{
\d^{m_im_j}T_{ \ldots m_i \ldots m_j \ldots} = 0\,,\quad\forall i,j\,.
}

\paragraph{Garnir symmetries} The manifest symmetries in \tabtotensor\ do not describe all the symmetries of the tensor. In
general, tensors obey additional symmetries described by the Garnir relations among
Young tableaux \refs{\gdjames,\sagan,\stahngarnir}. Suppose column $j$ of the Young diagram
mapped to a tensor $T$ has
$b_j$ boxes and let $M=m_1m_2 \ldots m_p$ and $N=n_1n_2 \ldots n_q$ be indices from columns $j$
and $j+1$. If $p+q> b_j$ then
\eqn\garnirdef{
T_{ \ldots[MN] \ldots} = 0\,.
}
\def\youngdim{0.5}
For example, the Garnir symmetries of the tensor associated to the Young diagram
$\yng(4,2,1,1)$ (see \tabtotensor) can be written as
\eqn\gEx{
T_{[12345]678}=0\,,\quad T_{1234[567]8}=0\,,\quad T_{1[23456]78}=0\,.
}
Note that if columns $j$ and $j+1$ have the same number $q$ of boxes, then it is also true that
swapping these two columns is a symmetry:
\eqn\swapb{
T_{ \ldots m_1 \ldots m_q n_1 \ldots n_q \ldots }= T_{ \ldots n_1 \ldots n_q m_1 \ldots m_q \ldots }
}
The symmetry \swapb\ is not independent of the symmetries \garnirdef. This is easier to see with the
alternative description of the Garnir symmetries reviewed in section~\altgarnirsec.

\newsubsec\oocsec open$\mkern1mu\otimes\mkern1mu$open=closed for first-level massive  states

The {\it massive} states of the superstring
combine to representations of $SO(9)$ but the amplitude calculations
are done in the Wick-rotated ten-dimensional spacetime, where the states are
described by $SO(10)$ irreps.

Consider all the $SO(10)$ irreps
in the tensor products \LiEprogram\ of $g_{mn}=(20000)$ and $b_{mnp}=(00100)$ appearing in the
open$\mkern1mu\otimes\mkern1mu$open
description of the closed-string states:
\eqnn\prods
$$\eqalignno{
(20000)\otimes(00100) &= (00100)+(10011)+(11000)+(20100) &\prods\cr
(20000)\otimes(20000) &= (00000)+(01000)+(02000)+(20000)+(21000)+(40000)\cr
(00100)\otimes(00100) &= (00000)+2(00011)+(00200)+(01000)+(01011)+(02000)\cr
&{}+(10002)+(10020)+(10100)+(20000)
}$$
To each irrep $(a_1a_2a_3a_4a_5)$ we
associate a transverse $SO(10)$ tensor
$g_{a_1a_2a_3a_4a_5}^{m \ldots}$. Their precise ranks and symmetries are determined from the
mapping to Young diagrams described above. For example, in $SO(10)$ we have
${\bf320}=(11000) \leftrightarrow \yng(2,1)$ and therefore the associated tensor $g^{mnp}_{11000}$ has three
indices with the symmetries of $\yng(2,1)$
\eqn\gooex{
k_m g_{11000}^{mnp} = 0,\quad g_{11000}^{mnp} = - g_{11000}^{nmp},\quad
g_{11000}^{mnp} + g_{11000}^{npm} + g_{11000}^{pmn} = 0\,.
}
The branching to $SO(9)$ irreps is given by $\bf 320 = \bf9+\bf36+\bf44+\bf231$ \LieArt. The
transverse and traceless conditions remove the
lower dimensional irreps and we are left with $\bf231$. In this sense we will call the tensor
$g_{11000}^{mnp}$ the $\bf231$ of $SO(9)$. This leads to the following decompositions:

\topinsert
\def\youngdim{0.4}
\def\tablerule{\noalign{\hrule}}
\centerline{\vbox{\offinterlineskip
\halign{\strut\hskip1pt#\hskip1pt\hfil
&\vrule\hfil\hskip1pt #\hfil
&\vrule\hfil\hskip1pt #\hfil
&\vrule\hfil\hskip1pt #\hfil
&\vrule\hfil\hskip1pt #\hfil
&\vrule\hfil\hskip1pt #\hfil
\vrule\cr
Dim & Dynkin  & Young & $44\otimes 44$ & $84\otimes 84$ & $84\otimes 44$ \cr\tablerule
1 & (0000)    & .            & $\surd$ & $\surd$ & $\times$  \cr
36 & (0100)   & \yng(1,1)    & $\surd$ & $\surd$ & $\times$  \cr
44 & (2000)   & \raise3pt\hbox{\yng(2)}      & $\surd$ & $\surd$ & $\times$  \cr
84 & (0010)   & \yng(1,1,1)  & $\times$ & $\epsilon_9$ & $\surd$  \cr
126 & (0002)  & \hskip-3pt\raise-1pt\hbox{\yng(1,1,1,1)}& $\times$ & $\surd$ & $\times$  \cr
231 & (1100)  & \yng(2,1)    & $\times$ & $\times$ & $\surd$  \cr
450 & (4000)  & \raise3pt\hbox{\yng(4)}      & $\surd$ & $\times$ & $\times$  \cr
495 & (0200)  & \raise2pt\hbox{\yng(2,2)}    & $\surd$ & $\surd$ & $\times$   \cr
594 & (1010)  & \yng(2,1,1)  & $\times$ & $\surd$ & $\times$  \cr
910 & (2100)  & \yng(3,1)    & $\surd$ & $\times$ & $\times$  \cr
924 & (1002)  & \yng(2,1,1,1)& $\times$ & $\epsilon_9$ & $\surd$   \cr
1980 & (0020) & \yng(2,2,2)  & $\times$ & $\surd$ & $\times$  \cr
2457 & (2010) & \yng(3,1,1)  & $\times$ & $\times$ & $\surd$  \cr
2772 & (0102) & \raise-1pt\hbox{\yng(2,2,1,1)}& $\times$ & $\surd$ & $\times$ \cr
}}}
\smallskip
{\leftskip=0pt\rightskip=20pt\baselineskip6pt\noindent\ninepoint
{\bf Table 1.} The first-level massive closed-string states characterized by their dimension,
$SO(9)$ Dynkin labels and Young diagram. The columns ${\rm dim}\otimes{\rm dim}$ indicate the presence
($\surd$) or absence ($\times$) of
the closed-string state in the tensor product of open-string states given in \tpirreps.
The entries $\e_9$ indicate that the state contains a 9-dimensional Levi-Civita tensor.
\par}
\endinsert

\def\youngdim{0.5}
\item{$\bullet$} The closed-string state $\bf1$:
\eqnn\expdec
$$\eqalignno{
        g_{mn}\otimes\tilde g_{pq} &= \Pi^{mn}_{pq}g_{00000} &\expdec\cr
        b_{mnp}\otimes\tilde b_{qrs} &= \ap\Pi^{mnp}_{qrs}g_{00000}\cr
}$$
\item{$\bullet$} The closed-string state ${\bf36}=\yng(1,1)$
\eqnn\expts
$$\eqalignno{
g_{mn}\otimes\tilde g_{pq} &= \Pi^{mn}_{am_1}\Pi^{pq}_{am_2}g_{01000}^{m_1m_2}&\expts\cr
b_{mnp}\otimes\tilde b_{qrs} &= \ap\Pi^{mnp}_{abm_1}\Pi^{pqr}_{abm_2}g_{01000}^{m_1m_2}\cr
}$$
\item{$\bullet$} The closed-string state ${\bf44}=\yng(2)$
\eqnn\expff
$$\eqalignno{
g_{mn}\otimes\tilde g_{pq} &= \Pi^{mn}_{am_1}\Pi^{pq}_{am_2}g_{20000}^{m_1m_2}&\expff\cr
b_{mnp}\otimes\tilde b_{qrs} &= {13\over2}\ap\Pi^{mnp}_{abm_1}\Pi^{pqr}_{abm_2}g_{20000}^{m_1m_2}\cr
}$$
\item{$\bullet$} The closed-string state ${\bf84}=\yng(1,1,1)$
\eqnn\expef
$$\eqalignno{
        g_{mn}\otimes\tilde b_{qrs} &=
        \tilde g_{mn}\otimes b_{qrs} = \Pi^{mn}_{am_1}\Pi^{qrs}_{am_2m_3}
	g_{00100}^{m_1m_2m_3}&\expef\cr
}$$
\item{$\bullet$} The closed-string state ${\bf126}=\yng(1,1,1,1)$
\eqnn\expots
$$\eqalignno{
        b_{mnp}\otimes\tilde b_{qrs} &=
	\ap\Pi^{mnp}_{am_1m_2}\Pi^{pqr}_{am_3m_4}g_{00011}^{m_1m_2m_3m_4}
	+ \hat\epsilon_{10}^{mnpqrsabcd}g_{00011}^{abcd}&\expots\cr
}$$
\item{$\bullet$} The closed-string state ${\bf231}=\yng(2,1)$
\eqnn\exptto
$$\eqalignno{
	g_{mn}\otimes\tilde b_{qrs} &= \Pi^{mn}_{am_1}\Pi^{qrs}_{am_2m_3} g_{11000}^{m_2m_3m_1}&\exptto\cr
}$$
\item{$\bullet$} The closed-string state ${\bf450}=\yng(4)$
\eqnn\expffz
$$\eqalignno{
        g_{mn}\otimes\tilde g_{pq} &= g_{40000}^{mnpq}&\expffz\cr
}$$
\item{$\bullet$} The closed-string state ${\bf495}=\yng(2,2)$
\eqnn\expfnf
$$\eqalignno{
	g_{mn}\otimes\tilde g_{pq} &=
        g_{02000}^{mpqn}+g_{02000}^{npqm}
        + g_{02000}^{mqpn}+g_{02000}^{nqpm}
	&\expfnf\cr
        b_{mnp}\otimes\tilde b_{qrs} &=-{27\over5}\ap\Pi^{mnp}_{am_1m_2}\Pi^{pqr}_{am_3m_4}g_{02000}^{m_1m_2m_3m_4}\cr
}$$
\item{$\bullet$} The closed-string state ${\bf594}=\yng(2,1,1)$
\eqnn\expfnfour
$$\eqalignno{
	b_{mnp}\otimes\tilde b_{qrs} &=
	\ap\Pi^{mnp}_{am_1m_2}\Pi^{pqr}_{am_3m_4}g_{10100}^{m_1m_2m_3m_4}&\expfnfour\cr
}$$
\item{$\bullet$} The closed-string state ${\bf910}=\yng(3,1)$
\eqnn\expnoz
$$\eqalignno{
        g_{mn}\otimes\tilde g_{pq} &=
        g_{21000}^{mpqn} + g_{21000}^{npqm}
        + g_{21000}^{mqpn} + g_{21000}^{nqpm}&\expnoz\cr
}$$
\item{$\bullet$} The closed-string state ${\bf924}=\yng(2,1,1,1)$
\eqnn\expntf
$$\eqalignno{
	g_{mn} \otimes\tilde b_{pqr} &= \tilde g_{mn} \otimes b_{pqr} =
        g_{10011}^{mpqrn}
        + g_{10011}^{npqrm}
        &\expntf\cr
}$$
\item{$\bullet$} The closed-string state ${\bf1980}=\yng(2,2,2)$
\eqnn\exponez
$$\eqalignno{
        b_{mnp}\otimes\tilde b_{abc} &= g_{00200}^{mnpabc}&\exponez\cr
}$$
\item{$\bullet$} The closed-string state ${\bf2457}=\yng(3,1,1)$
\eqnn\exptffs
$$\eqalignno{
g_{mn} \otimes\tilde b_{qrs} & = \tilde g_{mn} \otimes b_{qrs} = g_{20100}^{qrsmn}&\exptffs\cr
}$$
\item{$\bullet$} The closed-string state ${\bf2772}=\yng(2,2,1,1)$
\eqnn\exptsst
$$\eqalignno{
b_{mnp}\otimes\tilde b_{qrs} &= g_{01011}^{mnpqrs} + g_{01011}^{mnpsqr} + g_{01011}^{mnprsq}
&\exptsst
}$$
and we ignore the $SO(9)$ states associated with $(10002)$ and $(10020)$ as their $SO(10)$ Young
diagrams have columns with five boxes which imply that the $SO(9)$ states are proportional to a
nine-dimension Levi-Civita.
\def\youngdim{0.5}%
In the above the various tensors $g_{ \ldots}$ have the same symmetries of their associated
Young diagrams and
$\Pi^{mn}_{ab}$ and
$\Pi^{mnp}_{abc}$ are the Young projectors \hansenI\ (see also
\refs{\hansenII,\hansenIII,\robbins}):
\eqnn\projs
$$\eqalignno{
\Pi^{mn}_{pq}&=\half(\hat \d_{mp}\hat\d_{nq}+\hat
\d_{mq}\hat\d_{np})-{1\over9}\hat\d_{mn}\hat\d_{pq} &\projs\cr
\Pi^{mnp}_{qrs} & = \hat\delta^{[m}_q\hat\d^n_r\hat\d^{p]}_s
}$$
where
\eqn\deltahat{
\hat\d_{mn}=\d_{mn}-{k_m k_n\over (k\cdot k)}
}
satisfies $k^m\hat\d_{mn} = 0$ for the first-level massive condition $\ap k^2=-1$.
Both projectors \projs\ are traceless and transverse if $\ap k^2=-1$.
These conditions are necessary in order for the
above decompositions to be traceless and transverse, since $g_{mn}$ and $b_{mnp}$ are both
traceless and transverse (w.r.t $k_1$). Furthermore
\eqn\hateps{
\hat\epsilon_{10}^{mnpqrsabcd} = \epsilon_{10}^{mnpqrsabcd} - 10!\ap k_1^{[m}\epsilon_{10}^{npqrsabcd]k_1}
}
satisfies $k_1^m \hat\epsilon_{10}^{mnpqrsabcd} = 0$.
If we denote $g_{mn}$ and $b_{mnp}$ by their open-string $SO(9)$ irrep dimensions $\bf44$ and $\bf84$, the
decompositions above can be summarized by the following tensor products \LiEprogram
\eqnn\tpirreps
$$\eqalignno{
\bf44\otimes \bf84 &= \bf84\oplus\bf231\oplus\bf924\oplus\bf2457 &\tpirreps\cr
\bf44\otimes \bf44 &= \bf1\oplus\bf36\oplus\bf44\oplus\bf450\oplus\bf495\oplus\bf910 \cr
\bf84\otimes \bf84 &=
\bf1\oplus\bf36\oplus\bf44\oplus\bf84\oplus\bf126\oplus\bf495\oplus\bf594\oplus\bf924\oplus\bf1980\oplus\bf2772
}$$

\newsubsubsec\altgarnirsec Alternative description of the Garnir symmetries

In this subsection we give an alternative description of the Garnir symmetries
following \FultonYoungbook, in which all the signs are positive. More precisely,
for a given tableau,
summing over all ways of exchanging the top $k$ elements
of one column with $k$ elements of the preceding column while
preserving the vertical orders within each set of $k$ elements gives back the
original tableau.

As an illustration, consider the tableau in
\tabtotensor\ and its associated tensor $T_{12345678}$.
To find all its non-trivial symmetries, we start by
considering the $k=1$ relation
for the second column. We get the corresponding identity
\def\youngdim{1.4}%
$$
\young(1578,26,3,4)
\raise14pt\hbox{$=\;\;$}\young(5178,26,3,4)
\raise14pt\hbox{$+\;\;$}\young(1278,56,3,4)
\raise14pt\hbox{$+\;\;$}\young(1378,26,5,4)
\raise14pt\hbox{$+\;\;$}\young(1478,26,3,5)
$$
\eqn\symo{
T_{12345678}=T_{5234 16 78} + T_{1534 26 78} + T_{1254 36 78} + T_{1235 46 78}
}
Similarly, $k=1$ for the third column
yields
$$
\young(1578,26,3,4)
\raise14pt\hbox{$=\;\;$}\young(1758,26,3,4)
\raise14pt\hbox{$+\;\;$}\young(1568,27,3,4)
$$
\eqn\symt{
T_{1234 567 8} = T_{1234 765 8} + T_{1234 576 8}\,,
}
It is easy to see that $k=1$ for the fourth column leads to
the trivial symmetry $(78)$ that can be read off from the tableau itself.

Now we consider the $k=2$ relation with
the top two elements $\{5,6\}$ of the second column in \tabtotensor. This yields
$$
\young(1578,26,3,4)
\raise14pt\hbox{$=\;\;$}\young(5178,62,3,4)
\raise14pt\hbox{$+\;\;$}\young(5178,23,6,4)
\raise14pt\hbox{$+\;\;$}\young(5178,24,3,6)
\raise14pt\hbox{$+\;\;$}\young(1278,53,6,4)
\raise14pt\hbox{$+\;\;$}\young(1278,54,3,6)
\raise14pt\hbox{$+\;\;$}\young(1378,24,5,6)
$$
\eqn\settwoG{
T_{1234 56 78} = T_{56341278}+T_{52641378}+T_{52361478}+T_{15642378}+T_{15362478}+T_{12563478}\,.
}
\def\youngdim{0.5}
It is not difficult to show that the above symmetries are equivalent to\foot{
To show the third equation, start from \settwoG\ and use
\symo\ to rewrite $T_{5 \ldots}=\sum T_{1 \ldots}$.}
\eqn\gSym{
T_{[12345]678}=0\,,\quad T_{1234[567]8}=0\,,\quad T_{1[23456]78}=0\,.
}
In this description of the Garnir symmetries it is easy to see that
if two adjacent columns have the same number of boxes, then swapping the
corresponding columns in the Young tableau is a symmetry (choose $k$ to be the total number of
boxes).
\paragraph{Young projections} There is a way to experimentally verify the above Garnir symmetries.
To see this for the specific example in \tabtotensor, define the tensor as
\eqn\TYT{
T(1, \ldots,8) = W_{1 \ldots 8} + (1578)+(26)+[1234]+[56]
}
where $W_{ \ldots}$ denote words and
the (anti)symmetrizations must follow the specified order: first symmetrize over $1,5,7,8$, then
symmetrize over $2,6$, then antisymmetrize over $1,2,3,4$ and then antisymmetrize over
$5,6$. This yields $2304$ words in the right-hand side of \TYT, and the expression is available to
download in \massivewebsite. The
symmetries \gSym\ are then satisfied by \TYT.

\ninerm
\listrefs

\bye